\documentclass[10pt]{article}
\usepackage{amsxtra}
\usepackage{amssymb, amsmath}
\usepackage{amsthm}
\usepackage{hyperref}
\usepackage{multicol}

\usepackage[dvips]{graphicx}
\usepackage{latexsym}

\usepackage[T1]{fontenc} % se deixar só essa linha, é a fonte computer modern (com serifa) - para a maior parte das fontes, esta linha deve ser deixada descomentada
\usepackage{palatino} % com serifa

\begin{document}
\begin{center}
\textbf{\LARGE{Representations of Genetic Tables, Bimagic Squares, Hamming Distances and Shannon Entropy}}
\end{center}

\smallskip
\begin{center}
\textbf{\large{Inder J. Taneja}}\\
Departamento de Matem\'{a}tica\\
Universidade Federal de Santa Catarina\\
88.040-900 Florian\'{o}polis, SC, Brazil.\\
\textit{e-mail: ijtaneja@gmail.com\\
http://www.mtm.ufsc.br/$\sim $taneja}
\end{center}

\begin{abstract}
In this paper we have established relations of the genetic tables with magic and bimagic squares. Connections with Hamming distances, binomial coefficients are established. The idea of Gray code is applied. Shannon entropy of magic squares of order $4\times 4$, $8\times 8$ and $16\times 16$ are also calculated. Some comparison is also made. Symmetry among restriction enzymes having four letters is also studied.
\end{abstract}


\section{Introduction}

Genetic code is the set of rules by which information encoded in RNA/DNA is
translated into amino acid sequences in living cells. The bases for the
encoded information are nucleotides. There are four nucleotide bases for
RNA: Adenine, Uracil, Guanine, and Cytosine, which are labeled by $A$, $U$,
$G$ and $C$ respectively, (in DNA Uracil is replaced by Thymine ($T))$. In
canonical genetic code, codons are tri-nucleotide sequences such that each
triplet relates to an amino acid. For example, the codon CAG encodes the
amino acid Glutamine. Amino acids are the basic building blocks of proteins.
It stimulated interest of other researchers to study how genetic code was
translated into amino acids. There are 20 different amino acids (plus start
and stop codons), and since there are four nucleotide bases, $A$, $U$, $T$
and $C$, there are $4^{n}$ different combinations of bases, for a string of
length $n$. Therefore, $n=3$ is the smallest number of bases that could be
used to represent the 20 different amino acids. There is degeneracy between
the codons, i.e., more than one codon can represent the same amino acid;
however, two different amino acids cannot be represented by the same codon.
The following \textbf{CODON} table is well-known in the literature.
\begin{table}[htbp]\footnotesize
\begin{center}
\begin{tabular}{|l|l|l|l|l|l|}
\hline
&
T&
C&
A&
G&
 \\
\hline
\raisebox{-4.50ex}[0cm][0cm]{T}&
TTT (\textit{Phe})&
TCT (\textit{Ser})&
TAT (\textit{Tyr})&
TGT (\textit{Cys})&
T \\
\cline{2-6}
 &
TTC (\textit{Phe})&
TCC (\textit{Ser})&
TAC (\textit{Tyr)}&
TGC (\textit{Cys})&
C \\
\cline{2-6}
 &
TTA (\textit{Leu})&
TCA (\textit{Ser})&
TAA (\textit{Stop})&
TGA (\textit{Stop})&
A \\
\cline{2-6}
 &
TTG (\textit{Leu})&
TCG (\textit{Ser})&
TAG (\textit{Stop})&
TGG (\textit{Trp})&
G \\
\hline
\raisebox{-4.50ex}[0cm][0cm]{C}&
CTT (\textit{Leu})&
CCT (\textit{Pro})&
CAT (\textit{His})&
CGT (\textit{Arg})&
T \\
\cline{2-6}
 &
CTC (\textit{Leu})&
CCC (\textit{Pro})&
CAC (\textit{His})&
CGC (\textit{Arg})&
C \\
\cline{2-6}
 &
CTA (\textit{Leu})&
CCA (\textit{Pro})&
CAA (\textit{Glu})&
CGA (\textit{Arg})&
A \\
\cline{2-6}
 &
CTG (\textit{Leu})&
CCG (\textit{Pro})&
CAG (\textit{Glu})&
CGG (\textit{Arg})&
G \\
\hline
\raisebox{-4.50ex}[0cm][0cm]{A}&
ATT (\textit{Ile})&
ACT (\textit{Thr})&
AAT (\textit{Asn})&
AGT (\textit{Ser})&
T \\
\cline{2-6}
 &
ATC (\textit{Ile})&
ACC (\textit{Thr})&
AAC (\textit{Asn})&
AGC (\textit{Ser})&
C \\
\cline{2-6}
 &
ATA (\textit{Ile})&
ACA (\textit{Thr})&
AAA (\textit{Lys})&
AGA (\textit{Arg})&
A \\
\cline{2-6}
 &
ATG (\textit{Met})&
ACG (\textit{Thr})&
AAG (\textit{Lys})&
AGG (\textit{Arg})&
G \\
\hline
\raisebox{-4.50ex}[0cm][0cm]{G}&
GTT (\textit{Val})&
GCT (\textit{Ala})&
GAT (\textit{Asp})&
GGT (\textit{Gly})&
T \\
\cline{2-6}
 &
GTC (\textit{Val})&
GCC (\textit{Ala})&
GAC (\textit{Asp})&
GGC (\textit{Gly})&
C \\
\cline{2-6}
 &
GTA (\textit{Val})&
GCA (\textit{Ala})&
GAA (\textit{Glu})&
GGA (\textit{Gly})&
A \\
\cline{2-6}
 &
GTG (\textit{Val})&
GCG (\textit{Ala})&
GAG (\textit{Glu})&
GGG (\textit{Gly})&
G \\
\hline
\end{tabular}
\label{tab1}

\end{center}
\end{table}

The DNA (Deoxyribonucleic acid) molecule residing in the cell nucleus
encodes information conventionally represented as a symbolic string over the
alphabet. The combination between single strands of DNA takes place
according to ``Watson-Crick \cite{a12} complementarity'' that says that the only
permissible combinations between bases are $A-T$ or $T-A$ and $C-G$ or $G-C$
hence one strand can easily be used to predict the other in a double
stranded chain. Let us consider the following configurations of $4^{n}$ for
each value of $n$.

\begin{itemize}
\item [(i)] For $n=1$: In this case we have $4^{1}=4$. This gives
\[
M_{1} :=\left[ {{\begin{array}{*{20}c}
 C \hfill & A \hfill \\
 T \hfill & G \hfill \\
\end{array} }} \right].
\]

\item [(ii)] For $n=2$: In this case we have $4^{2}=16$. This gives
\[
M_{2} :=\left[ {{\begin{array}{*{20}c}
 {CC} \hfill & {AC} \hfill & {TC} \hfill & {GC} \hfill \\
 {CA} \hfill & {AA} \hfill & {TA} \hfill & {GA} \hfill \\
 {CT} \hfill & {AT} \hfill & {TT} \hfill & {GT} \hfill \\
 {CG} \hfill & {AG} \hfill & {TG} \hfill & {GG} \hfill \\
\end{array} }} \right].
\]

\item [(iii)] For $n=3$: In this case we have $4^{3}=64$. This gives
\[
M_{3} :=\left[ {{\begin{array}{*{20}c}
 {CCC} \hfill & {ACC} \hfill & {TCC} \hfill & {GCC} \hfill & {CTC} \hfill &
{ATC} \hfill & {TTC} \hfill & {GTC} \hfill \\
 {CCA} \hfill & {ACA} \hfill & {TCA} \hfill & {GCA} \hfill & {CTA} \hfill &
{ATA} \hfill & {TTA} \hfill & {GTA} \hfill \\
 {CCT} \hfill & {ACT} \hfill & {TCT} \hfill & {GCT} \hfill & {CTT} \hfill &
{ATT} \hfill & {TTT} \hfill & {GTT} \hfill \\
 {CCG} \hfill & {ACG} \hfill & {TCG} \hfill & {GCG} \hfill & {CTG} \hfill &
{ATG} \hfill & {TTG} \hfill & {GTG} \hfill \\
 {CAC} \hfill & {AAC} \hfill & {TAC} \hfill & {GAC} \hfill & {CGC} \hfill &
{AGC} \hfill & {TGC} \hfill & {GGC} \hfill \\
 {CAA} \hfill & {AAA} \hfill & {TAA} \hfill & {GAA} \hfill & {CGA} \hfill &
{AGA} \hfill & {TGA} \hfill & {GGA} \hfill \\
 {CAT} \hfill & {AAT} \hfill & {TAT} \hfill & {GAT} \hfill & {CGT} \hfill &
{AGT} \hfill & {TGT} \hfill & {GGT} \hfill \\
 {CAG} \hfill & {AAG} \hfill & {TAG} \hfill & {GAG} \hfill & {CGG} \hfill &
{AGG} \hfill & {TGG} \hfill & {GGG} \hfill \\
\end{array} }} \right].
\]

\item [(iv)] For $n=4$, we have $M_{4} $ with $4^{4}=256$ combinations of blocks of four letters, for $n=5$, we have $M_{5} $ with $4^{5}=1024$, etc.

\end{itemize}

\section{Gray Codes: Binary representations}

\subsection{First Approach}

Let us represent the letters $C$, $A$, $T$ and $G$ in two different ways:

\begin{itemize}
\item [(i)] $C=00,
\quad
A=01,
\quad
T=10$ and $G=11$.
\item [(ii)] $C=1,
\quad
A=2,
\quad
T=3$ and $G=4$.
\item [(iii)] The CODON table given above is formed of three letters out of four, i.e., $A$, $T$, $G$ and $C$. According to (i), we can write, for example, $TTA\sim 101001$, $AGC\sim 011100$, etc. Thus we have six digit binary representations of 64 members available in CODON table. Let us apply the change of base 2 to base 10 (decimal) using the formula $(abcdef)_{2} := a\cdot 2^{5}+b\cdot 2^{4}+c\cdot 2^{3}+d\cdot 2^{2}+e\cdot 2^{1}+f\cdot 2^{0}$ and then writing $(abcdef)_{2} +1$, we have, $TTA\sim 101001\sim 41$ and $AGC\sim 011100\sim 28$, etc. Similarly, we can write the four digits binary representation in decimal forms, such as $(abcd)_{2} :=a\cdot 2^{3}+b\cdot 2^{2}+c\cdot 2^{1}+d\cdot 2^{0}$, and then writing $(abcd)_{2} +1$. Just for simplicity, we have added 1.
\end{itemize}

The notations given in (i) and (ii) can be seen in \cite{a4, a6, a7, a8, a10}. Decimal
representation of the numbers is given by $C=\left( {00} \right)_{2} \sim
0$, $A=\left( {01} \right)_{2} \sim 1$, $T=\left( {10} \right)_{2} \sim 2$
and $G=\left( {11} \right)_{2} \sim 3$. For simplicity, we have added 1 and
considered in (ii) as 1, 2, 3, and 4 instead of 0, 1, 2 and 3 respectively.
We shall use frequently these three representations and shall bring magic
squares of different orders According to above notations we have
\begin{align}
\label{eq1}
M_{1} & :=\left[ {{\begin{array}{*{20}c}
 C \hfill & A \hfill \\
 T \hfill & G \hfill \\
\end{array} }} \right]\sim \left[ {{\begin{array}{*{20}c}
 {00} \hfill & {01} \hfill \\
 {10} \hfill & {11} \hfill \\
\end{array} }} \right]\sim \left[ {{\begin{array}{*{20}c}
 1 \hfill & 2 \hfill \\
 3 \hfill & 4 \hfill \\
\end{array} }} \right],\\
\label{eq2}
M_{2} & :=\left[ {{\begin{array}{*{20}c}
 {0000} \hfill & {0100} \hfill & {1000} \hfill & {1100} \hfill \\
 {0001} \hfill & {0101} \hfill & {1001} \hfill & {1101} \hfill \\
 {0010} \hfill & {0110} \hfill & {1010} \hfill & {1110} \hfill \\
 {0011} \hfill & {0111} \hfill & {1011} \hfill & {1111} \hfill \\
\end{array} }} \right],\\
\label{eq3}
M_{2} & :=\left[ {{\begin{array}{*{20}c}
 {11} \hfill & {21} \hfill & {31} \hfill & {41} \hfill \\
 {12} \hfill & {22} \hfill & {32} \hfill & {42} \hfill \\
 {13} \hfill & {23} \hfill & {33} \hfill & {43} \hfill \\
 {14} \hfill & {24} \hfill & {34} \hfill & {44} \hfill \\
\end{array} }} \right]
\intertext{and}
\label{eq4}
M_{2} & :=\left[ {{\begin{array}{*{20}c}
 1 \hfill & 5 \hfill & 9 \hfill & {13} \hfill \\
 2 \hfill & 6 \hfill & {10} \hfill & {14} \hfill \\
 3 \hfill & 7 \hfill & {11} \hfill & {15} \hfill \\
 4 \hfill & 8 \hfill & {12} \hfill & {16} \hfill \\
\end{array} }} \right].
\end{align}

The expressions appearing in (\ref{eq1}) are due to (i), (ii) and (iii). The
expression (\ref{eq2}) is due to (i), (\ref{eq3}) is due to (ii) and (\ref{eq3}) is due to
(iii). Similar tables can also be writen for the matrix $M_{3} $ Some of
them can seen in \cite{a4, a6, a7, a8, a10}.

\subsection{Second Approach}

Following \cite{a2, a3}, we use the following correspondence for the nucleotides and
two-bit Gray codes:
\[
C\sim \left( {{\begin{array}{*{20}c}
 0 \hfill \\
 0 \hfill \\
\end{array} }} \right),
\quad
A\sim \left( {{\begin{array}{*{20}c}
 0 \hfill \\
 1 \hfill \\
\end{array} }} \right),
\quad
T\sim \left( {{\begin{array}{*{20}c}
 1 \hfill \\
 0 \hfill \\
\end{array} }} \right) \,\mbox{and} \, G\sim \left( {{\begin{array}{*{20}c}
 1 \hfill \\
 1 \hfill \\
\end{array} }} \right).
\]
The genetic code-based matrix, which will contain all nucleotide strings of
length $n$ is defined as $M_{n} $. The Gray code sequences represented by
$M_{n} $ will be denoted by a $2^{n}\times 2^{n}$ matrix. Here are
corresponding Gray code representations
\[
M_{1} :=\left[ {{\begin{array}{*{20}c}
 C \hfill & A \hfill \\
 T \hfill & G \hfill \\
\end{array} }} \right]\sim \left[ {{\begin{array}{*{20}c}
 {\left( {\begin{array}{l}
 0 \\
 0 \\
 \end{array}} \right)} \hfill & {\left( {\begin{array}{l}
 0 \\
 1 \\
 \end{array}} \right)} \hfill \\
 {\left( {\begin{array}{l}
 1 \\
 0 \\
 \end{array}} \right)} \hfill & {\left( {\begin{array}{l}
 1 \\
 1 \\
 \end{array}} \right)} \hfill \\
\end{array} }} \right]\sim \left[ {{\begin{array}{*{20}c}
 0 \hfill & 1 \hfill \\
 1 \hfill & 0 \hfill \\
\end{array} }} \right].
\]
In information theory, the \textit{Hamming distance }between two strings of equal length is the number
of positions for which the corresponding symbols are different. Put another
way, it measures the minimum number of \textit{substitutions }required to change one into the
other, or the number of errors that transformed one string into the other.
Thus we observe that the \textit{Hamming distances} of letters $C$ and $G$ is 0 and of letters $A$ and
$T$is 1. Replacing the same in the other cases we have
\[
M_{2} :=\left[ {{\begin{array}{*{20}c}
 {CC} \hfill & {AC} \hfill & {TC} \hfill & {GC} \hfill \\
 {CA} \hfill & {AA} \hfill & {TA} \hfill & {GA} \hfill \\
 {CT} \hfill & {AT} \hfill & {TT} \hfill & {GT} \hfill \\
 {CG} \hfill & {AG} \hfill & {TG} \hfill & {GG} \hfill \\
\end{array} }} \right]\sim \left[ {{\begin{array}{*{20}c}
 0 \hfill & 1 \hfill & 1 \hfill & 0 \hfill \\
 1 \hfill & 2 \hfill & 2 \hfill & 1 \hfill \\
 1 \hfill & 2 \hfill & 2 \hfill & 1 \hfill \\
 0 \hfill & 1 \hfill & 1 \hfill & 0 \hfill \\
\end{array} }} \right]
\]
and
\[
M_{3} :=\left[ {{\begin{array}{*{20}c}
 0 \hfill & 1 \hfill & 1 \hfill & 1 \hfill & 1 \hfill & 2 \hfill & 2 \hfill
& 1 \hfill \\
 1 \hfill & 2 \hfill & 2 \hfill & 1 \hfill & 2 \hfill & 3 \hfill & 3 \hfill
& 2 \hfill \\
 1 \hfill & 2 \hfill & 2 \hfill & 1 \hfill & 2 \hfill & 3 \hfill & 3 \hfill
& 2 \hfill \\
 1 \hfill & 1 \hfill & 1 \hfill & 0 \hfill & 1 \hfill & 2 \hfill & 2 \hfill
& 1 \hfill \\
 1 \hfill & 2 \hfill & 2 \hfill & 1 \hfill & 0 \hfill & 1 \hfill & 1 \hfill
& 0 \hfill \\
 2 \hfill & 3 \hfill & 3 \hfill & 2 \hfill & 1 \hfill & 2 \hfill & 2 \hfill
& 1 \hfill \\
 2 \hfill & 3 \hfill & 3 \hfill & 2 \hfill & 1 \hfill & 2 \hfill & 2 \hfill
& 1 \hfill \\
 1 \hfill & 2 \hfill & 2 \hfill & 1 \hfill & 0 \hfill & 1 \hfill & 1 \hfill
& 0 \hfill \\
\end{array} }} \right].
\]
In the theory of discrete signal processing as a fundamental operation for
binary variables, modulo-2 addition is utilized broadly. By definition$, $the
modulo-2 addition of two numbers written in binary notation is made in a
bitwise manner in accordance with the following rules:
\[
0+0=0,\quad 1+0=1,\quad 0+1=1,\quad 1+1=0
\]
For example, modulo-2 addition of two binary numbers $110$ and $101$, gives
the result $110\oplus 101=011(3)$, where 3 is the decimal representation of
$011$. In case of $10$ and $01$, we have $10\oplus 01=11$(3), where 3 is the
decimal representation of $11(\oplus $ is the symbol for modulo-2
addition. The distance in this symmetry group is known as the \textit{Hamming distance}. The modulo-2
addition of any two binary numbers always results in a new number from the
same series. If any system of elements demonstrates its connection with
diadic shifts, it indicates that the structural organization of its system
is related to the logic of modulo-2 addition. In particular we have
\[
M_{1} :=\left[ {{\begin{array}{*{20}c}
 C \hfill & A \hfill \\
 T \hfill & G \hfill \\
\end{array} }} \right]\sim \left[ {{\begin{array}{*{20}c}
 {\left( {\begin{array}{l}
 0 \\
 0 \\
 \end{array}} \right)} \hfill & {\left( {\begin{array}{l}
 0 \\
 1 \\
 \end{array}} \right)} \hfill \\
 {\left( {\begin{array}{l}
 1 \\
 0 \\
 \end{array}} \right)} \hfill & {\left( {\begin{array}{l}
 1 \\
 1 \\
 \end{array}} \right)} \hfill \\
\end{array} }} \right]\sim \left[ {{\begin{array}{*{20}c}
 0 \hfill & 1 \hfill \\
 1 \hfill & 0 \hfill \\
\end{array} }} \right]
\]
\[
M_{2} :=\left[ {{\begin{array}{*{20}c}
 {CC} \hfill & {AC} \hfill & {TC} \hfill & {GC} \hfill \\
 {CA} \hfill & {AA} \hfill & {TA} \hfill & {GA} \hfill \\
 {CT} \hfill & {AT} \hfill & {TT} \hfill & {GT} \hfill \\
 {CG} \hfill & {AG} \hfill & {TG} \hfill & {GG} \hfill \\
\end{array} }} \right]\sim \left( {{\begin{array}{*{20}c}
 {00} \hfill & {10} \hfill & {10} \hfill & {00} \hfill \\
 {01} \hfill & {11} \hfill & {11} \hfill & {01} \hfill \\
 {01} \hfill & {11} \hfill & {11} \hfill & {01} \hfill \\
 {00} \hfill & {10} \hfill & {10} \hfill & {00} \hfill \\
\end{array} }} \right)
\]
and
\[
M_{3} :=\left[ {{\begin{array}{*{20}c}
 {000} \hfill & {100} \hfill & {100} \hfill & {000} \hfill & {010} \hfill &
{110} \hfill & {110} \hfill & {010} \hfill \\
 {001} \hfill & {101} \hfill & {101} \hfill & {001} \hfill & {011} \hfill &
{111} \hfill & {111} \hfill & {011} \hfill \\
 {001} \hfill & {101} \hfill & {101} \hfill & {001} \hfill & {011} \hfill &
{111} \hfill & {111} \hfill & {011} \hfill \\
 {000} \hfill & {100} \hfill & {100} \hfill & {000} \hfill & {010} \hfill &
{110} \hfill & {110} \hfill & {010} \hfill \\
 {010} \hfill & {110} \hfill & {110} \hfill & {010} \hfill & {000} \hfill &
{100} \hfill & {100} \hfill & {000} \hfill \\
 {011} \hfill & {111} \hfill & {111} \hfill & {011} \hfill & {001} \hfill &
{101} \hfill & {101} \hfill & {001} \hfill \\
 {011} \hfill & {111} \hfill & {111} \hfill & {011} \hfill & {001} \hfill &
{101} \hfill & {101} \hfill & {001} \hfill \\
 {010} \hfill & {110} \hfill & {110} \hfill & {010} \hfill & {000} \hfill &
{100} \hfill & {100} \hfill & {000} \hfill \\
\end{array} }} \right].
\]
The results are obtained by using the binary operations given above for
example, $GT\sim \left( {\begin{array}{l}
 11 \\
 10 \\
 \end{array}} \right)\sim 01$, i.e., $11\oplus 10=01$, $ACT\sim \left(
{\begin{array}{l}
 001 \\
 100 \\
 \end{array}} \right)\sim 101$, i.e., $001\oplus 100=101$, etc. The decimal
transformations are
\[
\left( {00} \right)_{2} \sim 0,
\quad
\left( {01} \right)_{2} \sim 1,
\quad
\left( {10} \right)_{2} \sim 2,
\quad
\left( {11} \right)_{2} \sim 3
\]
and
\[
\left( {000} \right)_{2} \sim 0,
\quad
\left( {001} \right)_{2} \sim 1,
\quad
\left( {010} \right)_{2} \sim 2,
\quad
\left( {011} \right)_{2} \sim 3
\]
\[
\left( {100} \right)_{2} \sim 4
\quad
\left( {101} \right)_{2} \sim 5,
\quad
\left( {110} \right)_{2} \sim 6,
\quad
\left( {111} \right)_{2} \sim 7
\]
This gives
\[
M_{2} :=\left[ {{\begin{array}{*{20}c}
 0 \hfill & 2 \hfill & 2 \hfill & 0 \hfill \\
 1 \hfill & 3 \hfill & 3 \hfill & 1 \hfill \\
 1 \hfill & 3 \hfill & 3 \hfill & 1 \hfill \\
 0 \hfill & 2 \hfill & 2 \hfill & 0 \hfill \\
\end{array} }} \right]
\]
and
\[
M_{3} :=\left[ {{\begin{array}{*{20}c}
 0 \hfill & 4 \hfill & 4 \hfill & 0 \hfill & 2 \hfill & 6 \hfill & 6 \hfill
& 2 \hfill \\
 1 \hfill & 5 \hfill & 5 \hfill & 1 \hfill & 3 \hfill & 7 \hfill & 7 \hfill
& 3 \hfill \\
 1 \hfill & 5 \hfill & 5 \hfill & 1 \hfill & 3 \hfill & 7 \hfill & 7 \hfill
& 3 \hfill \\
 0 \hfill & 4 \hfill & 4 \hfill & 0 \hfill & 2 \hfill & 6 \hfill & 6 \hfill
& 2 \hfill \\
 2 \hfill & 6 \hfill & 6 \hfill & 2 \hfill & 0 \hfill & 4 \hfill & 4 \hfill
& 0 \hfill \\
 3 \hfill & 7 \hfill & 7 \hfill & 3 \hfill & 1 \hfill & 5 \hfill & 5 \hfill
& 1 \hfill \\
 3 \hfill & 7 \hfill & 7 \hfill & 3 \hfill & 1 \hfill & 5 \hfill & 5 \hfill
& 1 \hfill \\
 2 \hfill & 6 \hfill & 6 \hfill & 2 \hfill & 0 \hfill & 4 \hfill & 4 \hfill
& 0 \hfill \\
\end{array} }} \right].
\]

\section{Reconfiguration Tables and Magic Squares}

This section deals with the reconfigurations of matrizes given above. These
reconfigurations are made in such a way that using the notations given in
section 2.1, lead use to magic squares or bimagic squares. Here below are
the definitions of magic and bimagic squares

\begin{itemize}
\item[\textbullet] A \textbf{magic square} is a collection of numbers put as a square matrix, where the sum of element of each row, each column and two principal diagonals are the same sum. For simplicity, let us write it as \textbf{S1}
\item[\textbullet] \textbf{Bimagic square} is a magic square where the sum of squares of each element of rows, columns and two principal diagonals are the same. For simplicity, let us write it as \textbf{S2}.
\end{itemize}

\subsection{Reconfiguration Tables of order 4x4}
	
Let us reconsider the matrix $M_{2} $ as
\begin{equation}
\label{eq5}
M_{2}^{4\times 4} :=\left[ {{\begin{array}{*{20}c}
 {AT} \hfill & {TG} \hfill & {CC} \hfill & {GA} \hfill \\
 {CA} \hfill & {GC} \hfill & {AG} \hfill & {TT} \hfill \\
 {GG} \hfill & {CT} \hfill & {TA} \hfill & {AC} \hfill \\
 {TC} \hfill & {AA} \hfill & {GT} \hfill & {CG} \hfill \\
\end{array} }} \right].
\end{equation}

In the above configuration, we have permutations of letters $C,\;A,\;T\;$ and
$G$ are the \textit{first} and \textit{second} places, in various situations, for example, in each row, in
each column, main diagonals, each group of order $2\times 2$, middle group
of order $2\times 2$, four corner elements, symmetrical diagonals, etc.
These configurations are of the following type
\[
\left[ {{\begin{array}{*{20}c}
 {AT} \hfill & {TG} \hfill \\
 {CA} \hfill & {GC} \hfill \\
\end{array} }} \right],
\quad
\left[ {{\begin{array}{*{20}c}
 {GC} \hfill & {AG} \hfill \\
 {CT} \hfill & {TA} \hfill \\
\end{array} }} \right],
\quad
\left[ {{\begin{array}{*{20}c}
 {CA} \hfill & {GC} \hfill & {AG} \hfill & {TT} \hfill \\
\end{array} }} \right],
\quad
\left[ {{\begin{array}{*{20}c}
 {CC} \hfill \\
 {AG} \hfill \\
 {TA} \hfill \\
 {GT} \hfill \\
\end{array} }} \right], \mbox{etc.}
\]

Here below are 20 combinations where the first and second members are the
permutations of the letters $C,\;A,\;T\;$ and $G$:
\begin{table}[htbp]\small
\begin{center}
\begin{tabular}{||l|l|l|l||l|l|l|l||l|l|l|l||l|l|l|l||l|l|l|l||}
\hline \hline
1&
2&
3&
4&
5&
5&
5&
5&
9&
9&
10&
10&
14&
15&
15&
14&
17&
19&
20&
18 \\
\hline
1&
2&
3&
4&
6&
6&
6&
6&
9&
9&
10&
10&
16&
13&
13&
16&
19&
17&
18&
20 \\
\hline
1&
2&
3&
4&
7&
7&
7&
7&
11&
11&
12&
12&
16&
13&
13&
16&
20&
18&
17&
19 \\
\hline
1&
2&
3&
4&
8&
8&
8&
8&
11&
11&
12&
12&
14&
15&
15&
14&
18&
20&
19&
17 \\
\hline \hline
\end{tabular}
\label{tab2}
\end{center}
\end{table}

According to notations (i), (ii) and (iii) given in section 2.1, we have

\begin{table}[htbp]\small
\begin{center}
\begin{tabular}{|p{20pt}|p{20pt}|p{20pt}|p{20pt}|}
\hline
0110 \par 23 \par 7&
1011 \par 34 \par 12&
0000 \par 11 \par 1&
1101 \par 42 \par 14 \\
\hline
0001 \par 12 \par 2&
1100 \par 41 \par 13&
0111 \newline
24 \par 8&
1010 \par 33 \par 11 \\
\hline
1111 \par 44 \par 16&
0010 \par 13 \par 3&
1001 \par 32 \par 10&
0100 \par 21 \par 5 \\
\hline
1000 \par 31 \par 9&
0101 \par 22 \par 6&
1110 \par 43 \par 15&
0011 \par 14 \par 4 \\
\hline
\end{tabular}
\label{tab3}
\end{center}
\end{table}

\newpage
In all the three situations we have magic squares of order $4\times 4$ with
$S1^{4\times 4}:=2222$, $S1^{4\times 4}:=110$ and $S1^{4\times 4}:=34$
respectively. The last one is well-known \textbf{\textit{Khajurao magic
square of order }}$4\times 4$:

\begin{table}[htbp]
\begin{center}
\begin{tabular}{|l|l|l|l|}
\hline
7&
12&
1&
14 \\
\hline
2&
13&
8&
11 \\
\hline
16&
3&
10&
5 \\
\hline
9&
6&
15&
4 \\
\hline
\end{tabular}
\label{tab4}
\end{center}
\end{table}

The above magic square of order $4\times 4$ is one of the
\textbf{\textit{most perfect magic}} square known in the literature. Its
connections with genetic code can be seen in [11]. This is one of the very
little work available on magic squares connecting DNA.

According to binary operations given in section 2.2, we have
\[
M_{B}^{4\times 4} :=\left[ {{\begin{array}{*{20}c}
 {11} \hfill & {10} \hfill & {00} \hfill & {01} \hfill \\
 {01} \hfill & {00} \hfill & {10} \hfill & {11} \hfill \\
 {00} \hfill & {01} \hfill & {11} \hfill & {10} \hfill \\
 {10} \hfill & {11} \hfill & {01} \hfill & {00} \hfill \\
\end{array} }} \right]\sim \left[ {{\begin{array}{*{20}c}
 3 \hfill & 2 \hfill & 0 \hfill & 1 \hfill \\
 1 \hfill & 0 \hfill & 2 \hfill & 3 \hfill \\
 0 \hfill & 1 \hfill & 3 \hfill & 2 \hfill \\
 2 \hfill & 3 \hfill & 1 \hfill & 0 \hfill \\
\end{array} }} \right].
\]
We observe that the matrix $M_{2}^{4\times 4} $ is a composition of two
\textbf{\textit{mutually orthogonal diagonalize Latin squares}}, while the
matrix $M_{B}^{4\times 4} $ is not a diagonalize Latin square.

\subsection{Reconfiguration Tables of order 8x8}

Here we shall reorganize the CODON table or the matrix $M_{3}$ given
in section 1 in such way that it becomes as magic square of order $8\times
8$. We shall present two different ways:
\begin{enumerate}
\item Four \textit{magic squares} of order $4\times 4$ of the sum $S1^{4\times 4}$ having all the properties of the configuration matrix $M_{2}^{4\times 4} $ given by (\ref{eq5}).
\item \textit{Bimagic square} of order $8\times 8$.
\end{enumerate}

\subsubsection{First Representations}

Let us consider the following reorganization of matrix $M_{3} $ or the above
CODON table:
\begin{table}[htbp]
\begin{center}
\begin{tabular}{||l|l|l|l||l|l|l|l||}
\hline \hline
CCC&
TAT&
GTG&
AGA&
CAA&
TCG&
GGT&
ATC \\
\hline
GTA&
AGG&
CCT&
TAC&
GGC&
ATT&
CAG&
TCA \\
\hline
AGT&
GTC&
TAA&
CCG&
ATG&
GGA&
TCC&
CAT \\
\hline
TAG&
CCA&
AGC&
GTT&
TCT&
CAC&
ATA&
GGG \\
\hline \hline
CTG&
TGA&
GCC&
AAT&
CGT&
TTC&
GAA&
ACG \\
\hline
GCT&
AAC&
CTA&
TGG&
GAG&
ACA&
CGC&
TTT \\
\hline
AAA&
GCG&
TGT&
CTC&
ACC&
GAT&
TTG&
CGA \\
\hline
TGC&
CTT&
AAG&
GCA&
TTA&
CGG&
ACT&
GAC \\
\hline \hline
\end{tabular}
\label{tab5}
\end{center}
\end{table}

In the above configuration we have the same properties of the magic square
of order $4\times 4$ given by (\ref{eq5}), i.e., there are many permutation of the
letters $C,\;A,\;T\;$ and $G$ in the \textit{first, second }and\textit{ third }places.
Each block of order $4\times 4$, half-row, half-column, half-principle diagonals etc. are also follow the
same property. There are much more combinations of this type in the above
configuration. In another way we can say there is a uniform distributions of
letters $C,\;A,\;T\;$ and $G$. Using the notations (i), (ii) and (iii) given
in section 2.1, we have
\newpage
\begin{table}[htbp]
\begin{center}
\begin{tabular}{||p{30pt}|p{30pt}|p{30pt}|p{30pt}||p{30pt}|p{30pt}|p{30pt}|p{30pt}||}
\hline \hline
000000 \par 111 \par 1&
100110 \par 323 \par 39&
111011 \par 434 \par 60&
011101 \par 242 \par 30&
000101 \par 122 \par 6&
100011 \par 314 \par 36&
111110 \par 443 \par 63&
011000 \par 231 \par 25 \\
\hline
111001 \par 432 \par 58&
011111 \par 244 \par 32&
000010 \par 113 \par 3&
100100 \par 321 \par 37&
111100 \par 441 \par 61&
011010 \par 233 \par 27&
000111 \par 124 \par 8&
100001 \par 312 \par 34 \\
\hline
011110 \par 243 \par 31&
111000 \par 431 \par 57&
100101 \par 322 \par 38&
000011 \par 114 \par 4&
011011 \par 234 \par 28&
111101 \par 442 \par 62&
100000 \par 311 \par 33&
000110 \par 123 \par 7 \\
\hline
100111 \par 324 \par 40&
000001 \par 112 \par 2&
011100 \par 241 \par 29&
111010 \par 433 \par 59&
100010 \par 313 \par 35&
000100 \par 121 \par 5&
011001 \par 232 \par 26&
111111 \par 444 \par 64 \\
\hline \hline
001011 \par 134 \par 12&
101101 \par 342 \par 46&
110000 \par 411 \par 49&
010110 \par 223 \par 23&
001110 \par 143 \par 15&
101000 \par 331 \par 41&
110101 \par 422 \par 54&
010011 \par 214 \par 20 \\
\hline
110010 \par 413 \par 51&
010100 \par 221 \par 21&
001001 \par 132 \par 10&
101111 \par 344 \par 48&
110111 \par 424 \par 56&
010001 \par 212 \par 18&
001100 \par 141 \par 13&
101010 \par 333 \par 43 \\
\hline
010101 \par 222 \par 22&
110011 \par 414 \par 52&
101110 \par 343 \par 47&
001000 \par 131 \par 9&
010000 \par 211 \par 17&
110110 \par 423 \par 55&
101011 \par 334 \par 44&
001101 \par 142 \par 14 \\
\hline
101100 \par 341 \par 45&
001010 \par 133 \par 11&
010111 \par 224 \par 24&
110001 \par 412 \par 50&
101001 \par 332 \par 42&
001111 \par 144 \par 16&
010010 \par 213 \par 19&
110100 \par 421 \par 53 \\
\hline \hline
\end{tabular}
\label{tab6}
\end{center}
\end{table}

The above table brings three different \textit{magic squares of order }$8\times 8, $i.e., in each case we have
$S1^{8\times 8}:=444444$, $S1^{8\times 8}:=2220$ and $S1^{8\times 8}:=260$
respectively. Moreover, the above magic square is also bimagic in columns,
i.e., for each column we have $S2^{8\times 8}:=\mbox{44893328844}$,
$S2^{8\times 8}:=\mbox{717}0\mbox{6}0$ and $S2^{8\times 8}:=\mbox{1118}0$
respectively. Also, each block of order $4\times 4$ is a magic square with
$S1^{4\times 4}:=222222$. Sum of each block of order $2\times 2$ also has
the same sum as of $S1^{4\times 4}$

\subsubsection{Second Representation}

We observe that the above magic square is bimagic only in columns. Here
below we shall present a little different representation of CODON table
resulting in \textit{bimagic square} of order $8\times 8$. Let us consider the following
configuration:
\begin{table}[htbp]
\begin{center}
\begin{tabular}{||l|l|l|l||l|l|l|l||}
\hline \hline
CGG&
TTC&
TCG&
CAC&
ATT&
GGA&
GAT&
ACA \\
\hline
ATA&
GGT&
GAA&
ACT&
CGC&
TTG&
TCC&
CAG \\
\hline \hline
CCC&
TAG&
TGC&
CTG&
AAA&
GCT&
GTA&
AGT \\
\hline
AAT&
GCA&
GTT&
AGA&
CCG&
TAC&
TGG&
CTC \\
\hline \hline
TAA&
CCT&
CTA&
TGT&
GCC&
AAG&
AGC&
GTG \\
\hline
GCG&
AAC&
AGG&
GTC&
TAT&
CCA&
CTT&
TGA \\
\hline \hline
TTT&
CGA&
CAT&
TCA&
GGG&
ATC&
ACG&
GAC \\
\hline
GGC&
ATG&
ACC&
GAG&
TTA&
CGT&
CAA&
TCT \\
\hline \hline
\end{tabular}
\label{tab7}
\end{center}
\end{table}

In the above configuration we have permutations of four letters
$C,\;A,\;T\;$ and $G$ only in the \textit{first} and \textit{third} place in each block of order 2x2,
half-row, half-column, half-principal diagonals etc. Again, using the
notations (i), (ii) and (iii) given in section 2.1, we have
\newpage
\begin{table}[htbp]
\begin{center}
\begin{tabular}{||p{30pt}|p{30pt}|p{30pt}|p{30pt}||p{30pt}|p{30pt}|p{30pt}|p{30pt}||}
\hline \hline
001111 \par 144 \par 16&
101000 \par 331 \par 41&
100011 \par 314 \par 36&
000100 \par 121 \par 5&
011010 \par 233 \par 27&
111101 \par 442 \par 62&
110110 \par 423 \par 55&
010001 \par 212 \par 18 \\
\hline
011001 \par 232 \par 26&
111110 \par 443 \par 63&
110101 \par 422 \par 54&
010010 \par 213 \par 19&
001100 \par 141 \par 13&
101011 \par 334 \par 44&
100000 \par 311 \par 33&
000111 \par 124 \par 8 \\
\hline \hline
000000 \par 111 \par 1&
100111 \par 324 \par 40&
101100 \par 341 \par 45&
001011 \par 134 \par 12&
010101 \par 222 \par 22&
110010 \par 413 \par 51&
111001 \par 432 \par 58&
011110 \par 243 \par 31 \\
\hline
010110 \par 223 \par 23&
110001 \par 412 \par 50&
111010 \par 433 \par 59&
011101 \par 242 \par 30&
000011 \par 114 \par 4&
100100 \par 321 \par 37&
101111 \par 344 \par 48&
001000 \par 131 \par 9 \\
\hline \hline
100101 \par 322 \par 38&
000010 \par 113 \par 3&
001001 \par 132 \par 10&
101110 \par 343 \par 47&
110000 \par 411 \par 49&
010111 \par 224 \par 24&
011100 \par 241 \par 29&
111011 \par 434 \par 60 \\
\hline
110011 \par 414 \par 52&
010100 \par 221 \par 21&
011111 \par 244 \par 32&
111000 \par 431 \par 57&
100110 \par 323 \par 39&
000001 \par 112 \par 2&
001010 \par 133 \par 11&
101101 \par 342 \par 46 \\
\hline \hline
101010 \par 333 \par 43&
001101 \par 142 \par 14&
000110 \par 123 \par 7&
100001 \par 312 \par 34&
111111 \par 444 \par 64&
011000 \par 231 \par 25&
010011 \par 214 \par 20&
110100 \par 421 \par 53 \\
\hline
111100 \par 441 \par 61&
011011 \par 234 \par 28&
010000 \par 211 \par 17&
110111 \par 424 \par 56&
101001 \par 332 \par 42&
001110 \par 143 \par 15&
000101 \par 122 \par 6&
100010 \par 313 \par 35 \\
\hline \hline
\end{tabular}
\label{tab8}
\end{center}
\end{table}

The above table bring us three \textbf{\textit{bimagic squares}} of order
8x8 with the following sums:

\begin{itemize}
\item [(a)] $S1^{8\times 8}:=444444=12012\times 37;
\quad
S2^{8\times 8}:=44893328844=\mbox{1213333212}\times \mbox{37}$
\item [(b)] $S1^{8\times 8}:=2220=60\times 37;
\quad
S2^{8\times 8}:=717060=19380\times 37$
\item [(c)] $S1^{8\times 8}:=260;
\quad
S2^{8\times 8}:=11180$.
\end{itemize}

We observe that (a) and (b) both $S1^{8\times 8}$ and $S2^{8\times 8}$ are
multiple of $37$. This is not true in case of (c). Still, in (a), (b) and
(c) each block of order $2\times 4$ is also \textit{bimagic}

\subsection{Connections with Prime Number 37}

Many authors \cite{a5, a6, a9} made connections with \textit{prime number} $37$. Here also we shall bring
some interesting connections with this prime number. Both the representations
we have three cases. In the first, we can write the sum $S1^{8\times
8}:=444444=12\times 1001\times 37$ In the second case we have $S1^{8\times
8}:=2220=12\times 5\times 37$. Also, in both these cases we have half-sum
of rows, columns and two principal diagonal as multiple of $37$, i.e., in
the first case we have $\textstyle{1 \over 2}S^{8\times 8}=222222=6\times
1001\times 37$, and in the second case we have $\textstyle{1 \over
2}S^{8\times 8}=1110=30\times 37$. There are many other combinations in the
above table giving connection with $37$. For example each block of $2\times
2$ is of sum $\textstyle{1 \over 2}S1^{8\times 8}$. In case of $S2^{8\times
8}$ we have $S2^{8\times 8}:=\mbox{44893328844}\mbox{=1213333212}\times
37$ and $S2^{8\times 8}:=\mbox{717}0\mbox{6}0=37\times 19380$. In
the first representation, we have \textit{bimagic} sum only in each column.

\subsection{Hamming Distances and Binomial Coefficients}

The idea of \textit{Hamming distances} is given in section 2.2. Here we consider more representations
to bring \textit{binomial coefficients} and and \textit{bimagic squares}. Kappraff and Adamson [4] considered $C=G$ and $A=U/T$. Following the idea of Watson and Crick [12], they [4] considered it as $A=T=2$ and $C=G=3$. For simplicity, let us consider here $A=T=a$ and $C=G=b$, where it is understood that $TTG=a\times a\times b=a^{2}b$, $AGC=a\times b\times b=ab^{2}$, etc. Accordingly, we have
\begin{itemize}
\item [(i)] For $n=1$:
\begin{table}[htbp]
\begin{center}
\begin{tabular}{|p{8pt}|p{8pt}|}
\hline
0 \par $b$&
1 \par $a$ \\
\hline
1 \par $a$&
0 \par $b$ \\
\hline
\end{tabular}
\label{tab9}
\end{center}
\end{table}

\item [(ii)]For $n=2$:
\begin{table}[htbp]
\begin{center}
\begin{tabular}{||p{12pt}|p{12pt}||p{12pt}|p{12pt}||}
\hline \hline
1 \par $ab$&
0 \par $b^{2}$&
2 \par $a^{2}$&
1 \par $ab$ \\
\hline
2 \par $a^{2}$&
1 \par $ab$&
1 \par $ab$&
0 \par $b^{2}$ \\
\hline \hline
0 \par $b^{2}$&
1 \par $ab$&
1 \par $ab$&
2 \par $a^{2}$ \\
\hline
1 \par $ab$&
2 \par $a^{2}$&
0 \par $b^{2}$&
1 \par $ab$ \\
\hline \hline
\end{tabular}
\label{tab10}
\end{center}
\end{table}

\item [(iii)]For $n=3$: According to configuration given in section 3.2.1, we have
\begin{table}[htbp]
\begin{center}
\begin{tabular}{||p{15pt}|p{15pt}|p{15pt}|p{15pt}||p{15pt}|p{15pt}|p{15pt}|p{15pt}||}
\hline \hline
0 \par $a^{3}$&
3 \par $b^{3}$&
1 \par $ab^{2}$&
2 \par $a^{2}b$&
2 \par $a^{2}b$&
1 \par $ab^{2}$&
1 \par $ab^{2}$&
2 \par $a^{2}b$ \\
\hline
2 \par $a^{2}b$&
1 \par $ab^{2}$&
1 \par $ab^{2}$&
2 \par $a^{2}b$&
0 \par $a^{3}$&
3 \par $b^{3}$&
1 \par $ab^{2}$&
2 \par $a^{2}b$ \\
\hline
2 \par $a^{2}b$&
1 \par $ab^{2}$&
3 \par $b^{3}$&
0 \par $a^{3}$&
2 \par $a^{2}b$&
1 \par $ab^{2}$&
1 \par $ab^{2}$&
2 \par $a^{2}b$ \\
\hline
2 \par $a^{2}b$&
1 \par $ab^{2}$&
1 \par $ab^{2}$&
2 \par $a^{2}b$&
2 \par $a^{2}b$&
1 \par $ab^{2}$&
3 \par $b^{3}$&
0 \par $a^{3}$ \\
\hline \hline
1 \par $ab^{2}$&
2 \par $a^{2}b$&
0 \par $a^{3}$&
3 \par $b^{3}$&
1 \par $ab^{2}$&
2 \par $a^{2}b$&
2 \par $a^{2}b$&
1 \par $ab^{2}$ \\
\hline
1 \par $ab^{2}$&
2 \par $a^{2}b$&
2 \par $a^{2}b$&
1 \par $ab^{2}$&
1 \par $ab^{2}$&
2 \par $a^{2}b$&
0 \par $a^{3}$&
3 \par $b^{3}$ \\
\hline
3 \par $b^{3}$&
0 \par $a^{3}$&
2 \par $a^{2}b$&
1 \par $ab^{2}$&
1 \par $ab^{2}$&
2 \par $a^{2}b$&
2 \par $a^{2}b$&
1 \par $ab^{2}$ \\
\hline
1 \par $ab^{2}$&
2 \par $a^{2}b$&
2 \par $a^{2}b$&
1 \par $ab^{2}$&
3 \par $b^{3}$&
0 \par $a^{3}$&
2 \par $a^{2}b$&
1 \par $ab^{2}$ \\
\hline \hline
\end{tabular}
\label{tab11}
\end{center}
\end{table}

According to configuration given in section 3.2.2 we have
\newpage
\begin{table}[htbp]
\begin{center}
\begin{tabular}{||p{15pt}|p{15pt}|p{15pt}|p{15pt}||p{15pt}|p{15pt}|p{15pt}|p{15pt}||}
\hline \hline
1 \par $ab^{2}$&
2 \par $a^{2}b$&
0 \par $b^{3}$&
1 \par $ab^{2}$&
2 \par $a^{2}b$&
3 \par $a^{3}$&
1 \par $ab^{2}$&
2 \par $ab^{2}$ \\
\hline
0 \par $b^{3}$&
1 \par $ab^{2}$&
1 \par $ab^{2}$&
2 \par $a^{2}b$&
1 \par $a^{2}b$&
2 \par $a^{2}b$&
2 \par $a^{2}b$&
3 \par $a^{3}$ \\
\hline \hline
0 \par $b^{3}$&
1 \par $ab^{2}$&
1 \par $ab^{2}$&
2 \par $a^{2}b$&
1 \par $ab^{2}$&
2 \par $a^{2}b$&
2 \par $a^{2}b$&
3 \par $a^{3}$ \\
\hline
1 \par $ab^{2}$&
2 \par $a^{2}b$&
0 \par $b^{3}$&
1 \par $ab^{2}$&
2 \par $a^{2}b$&
3 \par $a^{3}$&
1 \par $ab^{2}$&
2 \par $a^{2}b$ \\
\hline \hline
2 \par $a^{2}b$&
1 \par $ab^{2}$&
1 \par $ab^{2}$&
0 \par $b^{3}$&
3 \par $a^{3}$&
2 \par $a^{2}b$&
2 \par $a^{2}b$&
1 \par $ab^{2}$ \\
\hline
1 \par $ab^{2}$&
0 \par $b^{3}$&
2 \par $a^{2}b$&
1 \par $ab^{2}$&
2 \par $a^{2}b$&
1 \par $ab^{2}$&
3 \par $a^{3}$&
2 \par $a^{2}b$ \\
\hline \hline
1 \par $ab^{2}$&
0 \par $b^{3}$&
2 \par $a^{2}b$&
1 \par $ab^{2}$&
2 \par $a^{2}b$&
1 \par $ab^{2}$&
3 \par $a^{3}$&
2 \par $a^{2}b$ \\
\hline
2 \par $a^{2}b$&
1 \par $ab^{2}$&
1 \par $ab^{2}$&
0 \par $b^{3}$&
3 \par $a^{3}$&
2 \par $a^{2}b$&
2 \par $a^{2}b$&
1 \par $ab^{2}$ \\
\hline \hline
\end{tabular}
\label{tab12}
\end{center}
\end{table}
\end{itemize}

The interesting fact in the above tables is that in the first case, it is
symmetric in rows, columns and principal diagonals, while it is not true in
the second case. In the second case it holds only in rows. The tables
studied above gives us the following frequency distributions:
\begin{table}[htbp]
\begin{center}
\begin{tabular}{|p{10pt}|p{44pt}|p{58pt}|p{78pt}|p{36pt}|}
\hline
$n$&
Hamming \par distances&
Frequency \par distributions&
Binomial \par coefficients&
Sum \\
\hline
1&
0 1&
$2^{1}=2$&
$a\;\;b$&
$(a+b)^{1}$ \\
\hline
2&
0 1 2&
$2^{2}=4$&
$a^{2}\;\;2ab\;\;\;b^{2}$&
$(a+b)^{2}$ \\
\hline
3&
0 1 2 3&
$2^{3}=8$&
$a^{3}\;\;3a^{2}b\;\;3ab^{2}\;\;b^{2}$&
$(a+b)^{3}$ \\
\hline
\end{tabular}
\label{tab13}
\end{center}
\end{table}

For more properties of above table refer to \cite{a4}.

\subsection{Binary Operations}

Considering the notations and binary operations given in section 2.2, i.e.,
$CGT:=\left( {\begin{array}{l}
 011 \\
 010 \\
 \end{array}} \right)\sim 001$, $ATC:=\left( {\begin{array}{l}
 010 \\
 100 \\
 \end{array}} \right)\sim 110$, etc.  This operations gives us eight possibilities, i.e., $000$, $001$, $010$, $011$, $100$, $101$, $110$ and $111$. Let us represent $000\to a$, $001\to b$, $010\to c$, $011\to d$,
$100\to e$, $101\to f$, $110\to g$ and $111\to h$. Instead of decimal
representations as 0, 1, 2, 3, 4, 5, 6 and 7 we have considered here the
letters $a$, $b$, $c$, $d$, $e$, $f$, $g$ and $h$ respectively. According to
section 3.2.1, we have the following table:
\begin{table}[htbp]
\begin{center}
\begin{tabular}{||l|l||l|l||l|l||l|l||}
\hline \hline
$a$&
$h$&
$c$&
$f$&
$d$&
$e$&
$b$&
$g$ \\
\hline
$d$&
$e$&
$b$&
$g$&
$a$&
$h$&
$c$&
$f$ \\
\hline \hline
$f$&
$c$&
$h$&
$a$&
$g$&
$b$&
$e$&
$d$ \\
\hline
$g$&
$b$&
$e$&
$d$&
$f$&
$c$&
$h$&
$a$ \\
\hline \hline
$c$&
$f$&
$a$&
$h$&
$b$&
$g$&
$d$&
$e$ \\
\hline
$b$&
$g$&
$d$&
$e$&
$c$&
$f$&
$a$&
$h$ \\
\hline \hline
$h$&
$a$&
$f$&
$c$&
$e$&
$d$&
$g$&
$b$ \\
\hline
$e$&
$d$&
$g$&
$b$&
$h$&
$a$&
$f$&
$c$ \\
\hline \hline
\end{tabular}
\label{tab15}
\end{center}
\end{table}

We observe that we have 16 matrices of order $2\times 2$ divided in two
groups formed by the elements $(a,d,e,h)\sim (000,011,100,111)$ and
$(b,c,f,g)\sim (001,010,101,110)$. The above configuration is well-known
\textbf{\textit{diagonalize Latin square of order }}$8\times 8$ In the
second case, i.e., for the section 3.2.2, we don't have symmetric
configuration. See below:
\begin{table}[htbp]
\begin{center}
\begin{tabular}{||l|l||l|l||l|l||l|l||}
\hline \hline
$a$&
$g$&
$e$&
$c$&
$h$&
$b$&
$d$&
$f$ \\
\hline
$h$&
$b$&
$d$&
$f$&
$a$&
$g$&
$e$&
$c$ \\
\hline \hline
$a$&
$g$&
$e$&
$c$&
$h$&
$b$&
$d$&
$f$ \\
\hline
$h$&
$b$&
$d$&
$f$&
$a$&
$g$&
$e$&
$c$ \\
\hline \hline
$h$&
$b$&
$d$&
$f$&
$a$&
$g$&
$e$&
$c$ \\
\hline
$a$&
$g$&
$e$&
$c$&
$h$&
$b$&
$d$&
$f$ \\
\hline \hline
$h$&
$b$&
$d$&
$f$&
$a$&
$g$&
$e$&
$c$ \\
\hline
$a$&
$g$&
$e$&
$c$&
$h$&
$b$&
$d$&
$f$ \\
\hline \hline
\end{tabular}
\label{tab16}
\end{center}
\end{table}

\section{Restriction Enzymes}

There are (ref. Reiner [11]) 402 known restriction enzymes. Out of these 402
enzymes, 108 cut at a tetrameric sequence containing the four different
bases. Of these 108, 100{\%} have either $AT$ or $GC$ dimers (or both) in
the sequence. None of 108 enzymes have $G$ apart from $C$ and $A$ apart from
$T$, as in $AGTC$ Thus, all 108 enzymes cut at the tetrameric sequence which
is complementary to its reverse cyclic permutation. The specific
antiparallel sequences and the enzymes at which they cut are listed below.
Reiner [11] considered following two different combinations of four letters
having together, either $AT-TA$ or $GC-CG$ specifying antiparallel enzymes
See the table below:

\begin{table}[htbp]
\begin{center}
\begin{tabular}{||p{73pt}||p{73pt}||}
\hline \hline
Antiparallel \par $A-T,\;G-C$ \par in same \par orientation (88)&
Antiparallel \par $A-T,\;G-C$ \par in opposite \par orientation (20) \\
\hline \hline
$AGCT (9)$
&
$TAGC (0)$
 \\
\hline
$CGTA (0)$
&
$ACGT (2)$
 \\
\hline
$TACG (0)$
&
$GTAC (4)$
 \\
\hline
$CTAG (9)$
&
$GCTA (0)$
 \\
\hline
$GCAT (1)$
&
$TGCA (11)$
 \\
\hline
$TCGA (32)$
&
$ATCG (0)$
 \\
\hline
$ATGC (0)$
&
$CATG (3)$
 \\
\hline
$GATC (45)$
&
$CGAT (0)$
 \\
\hline \hline
\end{tabular}
\label{tab17}
\end{center}
\end{table}

Interestingly, the above pairs follows the same cyclic permutations, i.e.,
for example, $A-G-C-T-A-G-C$.

\subsection{Distribution of four Letter Combinations}

Very less work can be seen in literature having the combinations of \textit{four letters} in
\textit{four places}. Most of the work is towards \textit{codon representation} given above. Thus we observe that we
$4^{4}=256$ possibilities of writing combinations of four letters in four
places. Here below is a configuration $16\times 16$ having all the 256
possibilities.
\newpage
\begin{table}[htbp] \tiny
\begin{center}
\begin{tabular}{||l|l|l|l||l|l|l|l||l|l|l|l||l|l|l|l||}
\hline \hline
CCCC&
TATA&
GTGT&
AGAG&
CAAT&
TCGG&
GGTC&
ATCA&
CTTG&
TGCT&
GCAA&
AAGC&
CGGA&
TTAC&
GACG&
ACTT \\
\hline
GTAG&
AGGT&
CCTA&
TACC&
GGCA&
ATTC&
CAGG&
TCAT&
GCGC&
AAAA&
CTCT&
TGTG&
GATT&
ACCG&
CGAC&
TTGA \\
\hline
AGTA&
GTCC&
TAAG&
CCGT&
ATGG&
GGAT&
TCCA&
CATC&
AACT&
GCTG&
TGGC&
CTAA&
ACAC&
GAGA&
TTTT&
CGCG \\
\hline
TAGT&
CCAG&
AGCC&
GTTA&
TCTC&
CACA&
ATAT&
GGGG&
TGAA&
CTGC&
AATG&
GCCT&
TTCG&
CGTT&
ACGA&
GAAC \\
\hline \hline
CTGA&
TGAC&
GCCG&
AATT&
CGTG&
TTCT&
GAAA&
ACGC&
CCAT&
TAGG&
GTTC&
AGCA&
CACC&
TCTA&
GGGT&
ATAG \\
\hline
GCTT&
AACG&
CTAC&
TGGA&
GAGC&
ACAA&
CGCT&
TTTG&
GTCA&
AGTC&
CCGG&
TAAT&
GGAG&
ATGT&
CATA&
TCCC \\
\hline
AAAC&
GCGA&
TGTT&
CTCG&
ACCT&
GATG&
TTGC&
CGAA&
AGGG&
GTAT&
TACA&
CCTC&
ATTA&
GGCC&
TCAG&
CAGT \\
\hline
TGCG&
CTTT&
AAGA&
GCAC&
TTAA&
CGGC&
ACTG&
GACT&
TATC&
CCCA&
AGAT&
GTGG&
TCGT&
CAAG&
ATCC&
GGTA \\
\hline \hline
\underline {\textbf{\textit{CGAT}}}&
TTGG&
\underline {\textbf{\textit{GATC}}}&
ACCA&
CTCC&
TGTA&
GCGT&
AAAG&
CAGA&
TCAC&
GGCG&
ATTT&
CCTG&
TACT&
GTAA&
AGGC \\
\hline
GACA&
ACTC&
CGGG&
TTAT&
GCAG&
AAGT&
CTTA&
TGCC&
GGTT&
\underline {\textbf{\textit{ATCG}}}&
CAAC&
\underline {\textbf{\textit{TCGA}}}&
GTGC&
AGAA&
CCCT&
TATG \\
\hline
ACGG&
GAAT&
TTCA&
CGTC&
AATA&
GCCC&
TGAG&
CTGT&
ATAC&
GGGA&
TCTT&
CACG&
\underline {\textbf{\textit{AGCT}}}&
GTTG&
\underline {\textbf{\textit{TAGC}}}&
CCAA \\
\hline
TTTC&
CGCA&
ACAT&
GAGG&
TGGT&
\underline {\textbf{\textit{CTAG}}}&
AACC&
\underline {\textbf{\textit{GCTA}}}&
TCCG&
CATT&
ATGA&
GGAC&
TAAA&
CCGC&
AGTG&
GTCT \\
\hline \hline
\underline {\textbf{\textit{CATG}}}&
TCCT&
GGAA&
\underline {\textbf{\textit{ATGC}}}&
CCGA&
TAAC&
GTCG&
AGTT&
CGCC&
TTTA&
GAGT&
ACAG&
CTAT&
TGGG&
GCTC&
AACA \\
\hline
GGGC&
ATAA&
CACT&
TCTG&
GTTT&
AGCG&
CCAC&
TAGA&
GAAG&
\underline {\textbf{\textit{ACGT}}}&
\underline {\textbf{\textit{CGTA}}}&
TTCC&
GCCA&
AATC&
CTGG&
TGAT \\
\hline
ATCT&
GGTG&
TCGC&
CAAA&
AGAC&
GTGA&
TATT&
CCCG&
ACTA&
GACC&
TTAG&
CGGT&
AAGG&
\underline {\textbf{\textit{GCAT}}}&
\underline {\textbf{\textit{TGCA}}}&
CTTC \\
\hline
TCAA&
CAGC&
ATTG&
GGCT&
\underline {\textbf{\textit{TACG}}}&
CCTT&
AGGA&
\underline {\textbf{\textit{GTAC}}}&
TTGT&
CGAG&
ACCC&
GATA&
TGTC&
CTCA&
AAAT&
GCGG \\
\hline \hline
\end{tabular}
\label{tab18}
\end{center}
\end{table}

The construction of above table is based on the same techniques of the magic
of $M_{2}^{4\times 4} $. It has the e same properties as of $M_{2}^{4\times
4} $. Moreover, the \textbf{\textit{antiparallel pairs}} appearing in the
above table are in the same block in each case. They lies in the last eight
blocks of order $4\times 4$.

\subsection{Bimagic Squares of Order 16x16}

Let us consider now the representations (i) and (ii) of the letters
$C,\;A,\;T$ and $G$ as given in section 2.1. These representations lead us to
following two \textit{bimagic square of order }$16\times 16$.

\subsubsection{First representation}

This representation is according (i) given in section 2.1, by choosing
$C=00$, $A=01$, $T=10$ and $G=11$. This we have written in two parts:

\bigskip
\noindent \textbf{Part 1:}
\begin{table}[htbp]
\begin{center}
\begin{tabular}{||l|l|l|l||l|l|l|l||}
\hline \hline
00000000&
10011001&
11101110&
01110111&
00010110&
10001111&
11111000&
01100001 \\
\hline
11100111&
01111110&
00001001&
10010000&
11110001&
01101000&
00011111&
10000110 \\
\hline
01111001&
11100000&
10010111&
00001110&
01101111&
11110110&
10000001&
00011000 \\
\hline
10011110&
00000111&
01110000&
11101001&
10001000&
00010001&
01100110&
11111111 \\
\hline \hline
00101101&
10110100&
11000011&
01011010&
00111011&
10100010&
11010101&
01001100 \\
\hline
11001010&
01010011&
00100100&
10111101&
11011100&
01000101&
00110010&
10101011 \\
\hline
01010100&
11001101&
10111010&
00100011&
01000010&
11011011&
10101100&
00110101 \\
\hline
10110011&
00101010&
01011101&
11000100&
10100101&
00111100&
01001011&
11010010 \\
\hline \hline
00110110&
10101111&
11011000&
01000001&
00100000&
10111001&
11001110&
01010111 \\
\hline
11010001&
01001000&
00111111&
10100110&
11000111&
01011110&
00101001&
10110000 \\
\hline
01001111&
11010110&
10100001&
00111000&
01011001&
11000000&
10110111&
00101110 \\
\hline
10101000&
00110001&
01000110&
11011111&
10111110&
00100111&
01010000&
11001001 \\
\hline \hline
00011011&
10000010&
11110101&
01101100&
00001101&
10010100&
11100011&
01111010 \\
\hline
11111100&
01100101&
00010010&
10001011&
11101010&
01110011&
00000100&
10011101 \\
\hline
01100010&
11111011&
10001100&
00010101&
01110100&
11101101&
10011010&
00000011 \\
\hline
10000101&
00011100&
01101011&
11110010&
10010011&
00001010&
01111101&
11100100 \\
\hline \hline
\end{tabular}
\label{tab19}
\end{center}
\end{table}

\newpage
\noindent\textbf{Part 2:}
\begin{table}[htbp]
\begin{center}
\begin{tabular}{||l|l|l|l||l|l|l|l||}
\hline \hline
00101011&
10110010&
11000101&
01011100&
00111101&
10100100&
11010011&
01001010 \\
\hline
11001100&
01010101&
00100010&
10111011&
11011010&
01000011&
00110100&
10101101 \\
\hline
01010010&
11001011&
10111100&
00100101&
01000100&
11011101&
10101010&
00110011 \\
\hline
10110101&
00101100&
01011011&
11000010&
10100011&
00111010&
01001101&
11010100 \\
\hline \hline
00000110&
10011111&
11101000&
01110001&
00010000&
10001001&
11111110&
01100111 \\
\hline
11100001&
01111000&
00001111&
10010110&
11110111&
01101110&
00011001&
10000000 \\
\hline
01111111&
11100110&
10010001&
00001000&
01101001&
11110000&
10000111&
00011110 \\
\hline
10011000&
00000001&
01110110&
11101111&
10001110&
00010111&
01100000&
11111001 \\
\hline \hline
00011101&
10000100&
11110011&
01101010&
00001011&
10010010&
11100101&
01111100 \\
\hline
11111010&
01100011&
00010100&
10001101&
11101100&
01110101&
00000010&
10011011 \\
\hline
01100100&
11111101&
10001010&
00010011&
01110010&
11101011&
10011100&
00000101 \\
\hline
10000011&
00011010&
01101101&
11110100&
10010101&
00001100&
01111011&
11100010 \\
\hline \hline
00110000&
10101001&
11011110&
01000111&
00100110&
10111111&
11001000&
01010001 \\
\hline
11010111&
01001110&
00111001&
10100000&
11000001&
01011000&
00101111&
10110110 \\
\hline
01001001&
11010000&
10100111&
00111110&
01011111&
11000110&
10110001&
00101000 \\
\hline
10101110&
00110111&
01000000&
11011001&
10111000&
00100001&
01010110&
11001111 \\
\hline \hline
\end{tabular}
\label{tab20}
\end{center}
\end{table}

Let us combine the parts 1 and 2 as given below
\begin{table}[htbp]
\begin{center}
\begin{tabular}{|l|l|}
\hline
Part 1&
Part 2 \\
\hline
\end{tabular}
\label{tab21}
\end{center}
\end{table}

This gives us a bimagic square of order $16\times 16$ with $S1^{16\times
16}:=88888888$ and $S2^{16\times 16}:=\mbox{897867554657688}.$ Each block of
order $4\times 4$ is also a magic square with $S1^{4\times 4}:=22222222$.
Square of sum of each term in of each block of order $4\times 4$ is also
$S2^{16\times 16}:=\mbox{897867554657688}$ Here only $S2^{16\times 16}$ is
divisible by $37$ i.e, $S2^{16\times 16}:=\mbox{897867554657688}$
$=\mbox{2426669}0\mbox{666424}\times \mbox{37}$.

\subsubsection{Second representation}

This representation is according to (ii) given in section 2.1 by choosing
$C=1$, $A=1$, $T=3$ and $G=4$.
\begin{table}[htbp]\small
\begin{center}
\begin{tabular}{||l|l|l|l||l|l|l|l||l|l|l|l||l|l|l|l||}
\hline \hline
1111&
3232&
4343&
2424&
1223&
3144&
4431&
2312&
1334&
3413&
4122&
2241&
1442&
3321&
4214&
2133 \\
\hline
4324&
2443&
1132&
3211&
4412&
2331&
1244&
3123&
4141&
2222&
1313&
3434&
4233&
2114&
1421&
3342 \\
\hline
2432&
4311&
3224&
1143&
2344&
4423&
3112&
1231&
2213&
4134&
3441&
1322&
2121&
4242&
3333&
1414 \\
\hline
3243&
1124&
2411&
4332&
3131&
1212&
2323&
4444&
3422&
1341&
2234&
4113&
3314&
1433&
2142&
4221 \\
\hline \hline
1342&
3421&
4114&
2233&
1434&
3313&
4222&
2141&
1123&
3244&
4331&
2412&
1211&
3132&
4443&
2324 \\
\hline
4133&
2214&
1321&
3442&
4241&
2122&
1413&
3334&
4312&
2431&
1144&
3223&
4424&
2343&
1232&
3111 \\
\hline
2221&
4142&
3433&
1314&
2113&
4234&
3341&
1422&
2444&
4323&
3212&
1131&
2332&
4411&
3124&
1243 \\
\hline
3414&
1333&
2242&
4121&
3322&
1441&
2134&
4213&
3231&
1112&
2423&
4344&
3143&
1224&
2311&
4432 \\
\hline \hline
1423&
3344&
4231&
2112&
1311&
3432&
4143&
2224&
1242&
3121&
4414&
2333&
1134&
3213&
4322&
2441 \\
\hline
4212&
2131&
1444&
3323&
4124&
2243&
1332&
3411&
4433&
2314&
1221&
3142&
4341&
2422&
1113&
3234 \\
\hline
2144&
4223&
3312&
1431&
2232&
4111&
3424&
1343&
2321&
4442&
3133&
1214&
2413&
4334&
3241&
1122 \\
\hline
3331&
1412&
2123&
4244&
3443&
1324&
2211&
4132&
3114&
1233&
2342&
4421&
3222&
1141&
2434&
4313 \\
\hline \hline
1234&
3113&
4422&
2341&
1142&
3221&
4314&
2433&
1411&
3332&
4243&
2124&
1323&
3444&
4131&
2212 \\
\hline
4441&
2322&
1213&
3134&
4333&
2414&
1121&
3242&
4224&
2143&
1432&
3311&
4112&
2231&
1344&
3423 \\
\hline
2313&
4434&
3141&
1222&
2421&
4342&
3233&
1114&
2132&
4211&
3324&
1443&
2244&
4123&
3412&
1331 \\
\hline
3122&
1241&
2334&
4413&
3214&
1133&
2442&
4321&
3343&
1424&
2111&
4232&
3431&
1312&
2223&
4144 \\
\hline \hline
\end{tabular}
\label{tab20}
\end{center}
\end{table}

Here again we have bimagic square of order $16\times 16$ with $S1^{16\times
16}:=44440$ and $S2^{16\times 16}:=\mbox{14363412}0$. Each block of order
$4\times 4$ is also a magic square with $S1^{16\times 16}:=11110$. Square of
sum of each term in of each block of $4\times 4$ is also $S2^{16\times
16}:=\mbox{14363412}0$.

\subsubsection{Third representation}

Applying the change of base 2 to base 10 (decimal) in first case, i.e.,
\[
(abcdefgh)_{2} :=a\cdot 2^{7}+b\cdot 2^{6}+c\cdot 2^{5}+d\cdot 2^{4}+e\cdot
2^{3}+f\cdot 2^{2}+g\cdot 2^{1}+h\cdot 2^{0}
\]
then writing $(abcdefgh)_{2} +1$, we get the \textit{bimagic square of order }$16\times 16 $with sum
$S1^{16\times 16}:=2056$ and $S2^{16\times 16}:=351576$. Also each block of
order $4\times 4$ is a magic square with sum $S1^{4\times 4}:=514$.
\begin{table}[htbp]
\begin{center}
\begin{tabular}{||l|l|l|l||l|l|l|l||l|l|l|l||l|l|l|l||}
\hline \hline
1&
154&
239&
120&
23&
144&
249&
98&
44&
179&
198&
93&
62&
165&
212&
75 \\
\hline
232&
127&
10&
145&
242&
105&
32&
135&
205&
86&
35&
188&
219&
68&
53&
174 \\
\hline
122&
225&
152&
15&
112&
247&
130&
25&
83&
204&
189&
38&
69&
222&
171&
52 \\
\hline
159&
8&
113&
234&
137&
18&
103&
256&
182&
45&
92&
195&
164&
59&
78&
213 \\
\hline \hline
46&
181&
196&
91&
60&
163&
214&
77&
7&
160&
233&
114&
17&
138&
255&
104 \\
\hline
203&
84&
37&
190&
221&
70&
51&
172&
226&
121&
16&
151&
248&
111&
26&
129 \\
\hline
85&
206&
187&
36&
67&
220&
173&
54&
128&
231&
146&
9&
106&
241&
136&
31 \\
\hline
180&
43&
94&
197&
166&
61&
76&
211&
153&
2&
119&
240&
143&
24&
97&
250 \\
\hline \hline
55&
176&
217&
66&
33&
186&
207&
88&
30&
133&
244&
107&
12&
147&
230&
125 \\
\hline
210&
73&
64&
167&
200&
95&
42&
177&
251&
100&
21&
142&
237&
118&
3&
156 \\
\hline
80&
215&
162&
57&
90&
193&
184&
47&
101&
254&
139&
20&
115&
236&
157&
6 \\
\hline
169&
50&
71&
224&
191&
40&
81&
202&
132&
27&
110&
245&
150&
13&
124&
227 \\
\hline \hline
28&
131&
246&
109&
14&
149&
228&
123&
49&
170&
223&
72&
39&
192&
201&
82 \\
\hline
253&
102&
19&
140&
235&
116&
5&
158&
216&
79&
58&
161&
194&
89&
48&
183 \\
\hline
99&
252&
141&
22&
117&
238&
155&
4&
74&
209&
168&
63&
96&
199&
178&
41 \\
\hline
134&
29&
108&
243&
148&
11&
126&
229&
175&
56&
65&
218&
185&
34&
87&
208 \\
\hline \hline
\end{tabular}
\label{tab21}
\end{center}
\end{table}

According to above three constructions the Reiner \cite{a11} table of
antiparallelism is given by
\begin{table}[htbp] \small
\begin{center}
\begin{tabular}{||p{48pt}||p{32pt}|p{32pt}|p{32pt}|p{32pt}|p{32pt}|p{32pt}|p{32pt}|p{32pt}||p{32pt}||}
\hline \hline
Antiparallel \par A-T, G-C \par in same \par orientation&
AGCT \par 01110010 \par 2413 \par 115&
CGTA \par 00111001 \par 1432 \par 58&
TACG \par 10010011 \par 3214 \par 148&
CTAG \par 00100111 \par 1324 \par 40&
GCAT \par 11000110 \par 4123 \par 199&
TCGA \par 10001101 \par 3142 \par 142&
ATGC \par 01101100 \par 2341 \par 109&
GATC \par 11011000 \par 4231 \par 217&
\textbf{SUM} \par \textbf{44444444} \par \textbf{22220} \par \textbf{1028} \\
\hline \hline
Antiparallel \par A-T, G-C \par in opposite \par orientation&
TAGC \par 10011100 \par 3241 \par 157&
ACGT \par 01001110 \par 2143 \par 79&
GTAC \par 11100100 \par 4321 \par 229&
GCTA \par 11001001 \par 4132 \par 202&
TGCA \par 10110001 \par 3412 \par 178&
ATCG \par 01100011 \par 2314 \par 100&
CATG \par 00011011 \par 1234 \par 28&
CGAT \par 00110110 \par 1423 \par 55&
\textbf{SUM} \par \textbf{44444444} \par \textbf{22220} \par \textbf{1028} \\
\hline \hline
\end{tabular}
\label{tab22}
\end{center}
\end{table}

We have the same sum in both the situations, i.e., in the same as well as in
the opposite orientation of the genetic letters.

\subsection{Hamming Distances and Binomial Coefficients }

Here also we shall consider more representations to bring Hamming distances
and binomial coefficients. Using the same notations of section 3.4, we have
the following table with \textit{Hamming distances} and \textit{binomial coefficients}:
\begin{table}[htbp]\tiny
\begin{center}
\begin{tabular}{||p{15pt}|p{15pt}|p{15pt}|p{15pt}||p{15pt}|p{15pt}|p{15pt}|p{15pt}||p{15pt}|p{15pt}|p{15pt}|p{15pt}||p{15pt}|p{15pt}|p{15pt}|p{15pt}||}
\hline \hline
0 \par $b^{4}$&
4 \par $a^{4}$&
2 \par $a^{2}b^{2}$&
2 \par $a^{2}b^{2}$&
3 \par $a^{3}b$&
1 \par $ab^{3}$&
1 \par $ab^{3}$&
3 \par $a^{3}b$&
2 \par $a^{2}b^{2}$&
2 \par $a^{2}b^{2}$&
2 \par $a^{2}b^{2}$&
2 \par $a^{2}b^{2}$&
1 \par $ab^{3}$&
3 \par $a^{3}b$&
1 \par $ab^{3}$&
3 \par $a^{3}b$ \\
\hline
2 \par $a^{2}b^{2}$&
2 \par $a^{2}b^{2}$&
2 \par $a^{2}b^{2}$&
2 \par $a^{2}b^{2}$&
1 \par $ab^{3}$&
3 \par $a^{3}b$&
1 \par $ab^{3}$&
3 \par $a^{3}b$ &
0 \par $b^{4}$&
4 \par $a^{4}$&
2 \par $a^{2}b^{2}$&
2 \par $a^{2}b^{2}$&
3 \par $a^{3}b$&
1 \par $ab^{3}$&
1 \par $ab^{3}$&
3 \par $a^{3}b$ \\
\hline
3 \par $a^{3}b$&
1 \par $ab^{3}$&
3 \par $a^{3}b$&
1 \par $ab^{3}$&
2 \par $a^{2}b^{2}$&
2 \par $a^{2}b^{2}$&
2 \par $a^{2}b^{2}$&
2 \par $a^{2}b^{2}$&
3 \par $a^{3}b$&
1 \par $ab^{3}$&
1 \par $ab^{3}$&
3 \par $a^{3}b$&
2 \par $a^{2}b^{2}$&
2 \par $a^{2}b^{2}$&
4 \par $a^{4}$&
0 \par $b^{4}$ \\
\hline
3 \par $a^{3}b$&
1 \par $ab^{3}$&
1 \par $ab^{3}$&
3 \par $a^{3}b$&
2 \par $a^{2}b^{2}$&
2 \par $a^{2}b^{2}$&
4 \par $a^{4}$&
0 \par $b^{4}$&
3 \par $a^{3}b$&
1 \par $ab^{3}$&
3 \par $a^{3}b$&
1 \par $ab^{3}$&
2 \par $a^{2}b^{2}$&
2 \par $a^{2}b^{2}$&
2 \par $a^{2}b^{2}$&
2 \par $a^{2}b^{2}$ \\
\hline \hline
2 \par $a^{2}b^{2}$&
2 \par $a^{2}b^{2}$&
0 \par $b^{4}$&
4 \par $a^{4}$&
1 \par $ab^{3}$&
3 \par $a^{3}b$&
3 \par $a^{3}b$&
1 \par $ab^{3}$&
2 \par $a^{2}b^{2}$&
2 \par $a^{2}b^{2}$&
2 \par $a^{2}b^{2}$&
2 \par $a^{2}b^{2}$&
1 \par $ab^{3}$&
3 \par $a^{3}b$&
1 \par $ab^{3}$&
3 \par $a^{3}b$ \\
\hline
2 \par $a^{2}b^{2}$&
2 \par $a^{2}b^{2}$&
2 \par $a^{2}b^{2}$&
2 \par $a^{2}b^{2}$&
1 \par $ab^{3}$&
3 \par $a^{3}b$&
1 \par $ab^{3}$&
3 \par $a^{3}b$&
2 \par $a^{2}b^{2}$&
2 \par $a^{2}b^{2}$&
0 \par $b^{4}$&
4 \par $a^{4}$&
1 \par $ab^{3}$&
3 \par $a^{3}b$&
3 \par $a^{3}b$&
1 \par $ab^{3}$ \\
\hline
3 \par $a^{3}b$&
1 \par $ab^{3}$&
3 \par $a^{3}b$&
1 \par $ab^{3}$&
2 \par $a^{2}b^{2}$&
2 \par $a^{2}b^{2}$&
2 \par $a^{2}b^{2}$&
2 \par $a^{2}b^{2}$&
1 \par $ab^{3}$&
3 \par $a^{3}b$&
3 \par $a^{3}b$&
1 \par $ab^{3}$&
4 \par $a^{4}$ &
0 \par $b^{4}$&
2 \par $a^{2}b^{2}$&
2 \par $a^{2}b^{2}$ \\
\hline
1 \par $ab^{3}$&
3 \par $a^{3}b$&
3 \par $a^{3}b$&
1 \par $ab^{3}$&
4 \par $a^{4}$&
0 \par $b^{4}$&
2 \par $a^{2}b^{2}$&
2 \par $a^{2}b^{2}$&
3 \par $a^{3}b$&
1 \par $ab^{3}$&
3 \par $a^{3}b$&
1 \par $ab^{3}$&
2 \par $a^{2}b^{2}$&
2 \par $a^{2}b^{2}$&
2 \par $a^{2}b^{2}$&
2 \par $a^{2}b^{2}$ \\
\hline \hline
2 \par $a^{2}b^{2}$&
2 \par $a^{2}b^{2}$&
2 \par $a^{2}b^{2}$&
2 \par $a^{2}b^{2}$&
1 \par $ab^{3}$&
3 \par $a^{3}b$&
1 \par $ab^{3}$&
3 \par $a^{3}b$&
2 \par $a^{2}b^{2}$&
2 \par $a^{2}b^{2}$&
0 \par $b^{4}$&
4 \par $a^{4}$&
1 \par $ab^{3}$&
3 \par $a^{3}b$&
3 \par $a^{3}b$&
1 \par $ab^{3}$ \\
\hline
2 \par $a^{2}b^{2}$&
2 \par $a^{2}b^{2}$&
0 \par $b^{4}$&
4 \par $a^{4}$&
1 \par $ab^{3}$&
3 \par $a^{3}b$&
3 \par $a^{3}b$&
1 \par $ab^{3}$&
2 \par $a^{2}b^{2}$&
2 \par $a^{2}b^{2}$&
2 \par $a^{2}b^{2}$&
2 \par $a^{2}b^{2}$&
1 \par $ab^{3}$&
3 \par $a^{3}b$&
1 \par $ab^{3}$&
3 \par $a^{3}b$ \\
\hline
1 \par $ab^{3}$&
3 \par $a^{3}b$&
3 \par $a^{3}b$&
1 \par $ab^{3}$&
4 \par $a^{4}$&
0 \par $b^{4}$&
2 \par $a^{2}b^{2}$&
2 \par $a^{2}b^{2}$&
3 \par $a^{3}b$&
1 \par $ab^{3}$&
3 \par $a^{3}b$&
1 \par $ab^{3}$&
2 \par $a^{2}b^{2}$&
2 \par $a^{2}b^{2}$&
2 \par $a^{2}b^{2}$&
2 \par $a^{2}b^{2}$ \\
\hline
3 \par $a^{3}b$&
1 \par $ab^{3}$&
3 \par $a^{3}b$&
1 \par $ab^{3}$&
2 \par $a^{2}b^{2}$&
2 \par $a^{2}b^{2}$&
2 \par $a^{2}b^{2}$&
2 \par $a^{2}b^{2}$&
1 \par $ab^{3}$&
3 \par $a^{3}b$&
3 \par $a^{3}b$&
1 \par $ab^{3}$&
4 \par $a^{4}$&
0 \par $b^{4}$&
2 \par $a^{2}b^{2}$&
2 \par $a^{2}b^{2}$ \\
\hline \hline
2 \par $a^{2}b^{2}$&
2 \par $a^{2}b^{2}$&
2 \par $a^{2}b^{2}$&
2 \par $a^{2}b^{2}$&
1 \par $ab^{3}$&
3 \par $a^{3}b$&
1 \par $ab^{3}$&
3 \par $a^{3}b$&
0 \par $b^{4}$&
4 \par $a^{4}$&
2 \par $a^{2}b^{2}$&
2 \par $a^{2}b^{2}$&
3 \par $a^{3}b$&
1 \par $ab^{3}$&
1 \par $ab^{3}$&
3 \par $a^{3}b$ \\
\hline
0 \par $b^{4}$&
4 \par $a^{4}$&
2 \par $a^{2}b^{2}$&
2 \par $a^{2}b^{2}$&
3 \par $a^{3}b$&
1 \par $ab^{3}$&
1 \par $ab^{3}$&
3 \par $a^{3}b$&
2 \par $a^{2}b^{2}$&
2 \par $a^{2}b^{2}$&
2 \par $a^{2}b^{2}$&
2 \par $a^{2}b^{2}$&
1 \par $ab^{3}$&
3 \par $a^{3}b$&
1 \par $ab^{3}$&
3 \par $a^{3}b$ \\
\hline
3 \par $a^{3}b$&
1 \par $ab^{3}$&
1 \par $ab^{3}$&
3 \par $a^{3}b$&
2 \par $a^{2}b^{2}$&
2 \par $a^{2}b^{2}$&
4 \par $a^{4}$&
0 \par $b^{4}$&
3 \par $a^{3}b$&
1 \par $ab^{3}$&
3 \par $a^{3}b$&
1 \par $ab^{3}$&
2 \par $a^{2}b^{2}$&
2 \par $a^{2}b^{2}$&
2 \par $a^{2}b^{2}$&
2 \par $a^{2}b^{2}$ \\
\hline
3 \par $a^{3}b$&
1 \par $ab^{3}$&
3 \par $a^{3}b$&
1 \par $ab^{3}$&
2 \par $a^{2}b^{2}$&
2 \par $a^{2}b^{2}$&
2 \par $a^{2}b^{2}$&
2 \par $a^{2}b^{2}$&
3 \par $a^{3}b$&
1 \par $ab^{3}$&
1 \par $ab^{3}$&
3 \par $a^{3}b$&
2 \par $a^{2}b^{2}$&
2 \par $a^{2}b^{2}$&
4 \par $a^{4}$&
0 \par $b^{4}$ \\
\hline \hline
\end{tabular}
\label{tab28}
\end{center}
\end{table}

Here we observe the symmetry in elements in each row, each column and
each block of order $4\times 4$. The same symmetry we have in principal diagonals too.
This don't happened in case of order 8x8. Thus there is a straight relationship with the
binomial coefficients and Hamming distances, i.e.,
$0\to b^{4}$, $1\to ab^{3}$, $2\to a^{2}b^{2}$, $3\to a^{3}b$ and $4\to a^{4}$.
Accordingly, we have the following frequency distribution table:
\begin{itemize}
\item \textbf{Frequency distribution}
\end{itemize}
\begin{table}[htbp]
\begin{center}
\begin{tabular}{|p{10pt}|p{55pt}|p{55pt}|p{170pt}|p{36pt}|}
\hline
$n$&
Hamming \par distances&
Frequency \par distributions&
Binomial coefficients&
Sum \\
\hline
1&
0 1&
$2^{1}=2$&
$a\;b$&
$(a+b)^{1}$ \\
\hline
2&
0 1 2&
$2^{2}=4$&
$a^{2}\;2ab\;\;b^{2}$&
$(a+b)^{2}$ \\
\hline
3&
0 1 2 3&
$2^{3}=8$&
$a^{3}\;3a^{2}b\;3ab^{2}\;b^{2}$&
$(a+b)^{3}$ \\
\hline
4&
0 1 2 3 4&
$2^{4}=16$&
$a^{4}\;4a^{3}b\;6a^{2}b^{2}\;4ab^{3}\;a^{4}$&
$(a+b)^{4}$ \\
\hline
5&
0 1 2 3 4 5&
$2^{5}=32$&
$a^{5}\;5a^{4}b\;10a^{3}b^{2}\;10a^{2}b^{3}\;5ab^{4}\;b^{5}$&
$(a+b)^{5}$ \\
\hline
6&
0 1 2 3 4 5 6&
$2^{6}=64$&
$a^{6}\;6a^{5}b\;15a^{4}b^{2}\;20a^{3}b^{3}\;15a^{2}b^{4}\;6ab^{5}\;b^{6}$&
$(a+b)^{6}$ \\
\hline
\textellipsis &
\textellipsis &
\textellipsis &
\textellipsis &
\textellipsis \\
\hline
\end{tabular}
\label{tab29}
\end{center}
\end{table}

The above table allow us to extend the results for the next values of n.
Some studies having combinations of four letters can be seen in \cite{a1}.

\section{Shannon's Entropy and Genetic Tables}

The idea of Shannon entropy is well-known in the literature on information
theory. It is defined as
\[
H(P)=-\sum\limits_{i=1}^n {p_{i} \log p_{i} }
\]
where $P=(p_{1} ,p_{2} ,...,p_{n} )$, $p_{i} >0$, $\sum\limits_{i=1}^n
{p_{i} =1} $ is a set of probability distribution associated with a random
variable $X=\left\{ {x_{1} ,x_{2} ,...,x_{n} } \right\}$. Applications of
Shannon entroy to genetic code can be seen in many works In \cite{a13, a14},
authors introduce the idea of \textbf{\textit{genome order index}} given by
\[
S(P)=\sum\limits_{i=1}^n {p_{i}^{2} }
\]
In Information theory the expression $S(P)$ is famous as quadratic entropy.
Thus based the magic squares given above we shall calculate \textit{Shannon entropy} and \textit{genome order index} First, we shall transform values in probabilities dividing by sum of each row or column. Here we shall consider only the first case.

\subsection{Shannon Entropy of Order 4x4}

In section 3.1, we have three different kind of magic squares of order $4 \times 4$. The first one is with binary digits. In this let us divide the each value by the magic sum. This gives us the following probability distributions:
\begin{itemize}
\item\textbf{Probability distributions}
\end{itemize}

\begin{table}[htbp]
\begin{center}
\begin{tabular}{|l|l|l|l|}
\hline
0,4500&
0,0050&
0,0455&
0,4995 \\
\hline
0,0495&
0,4955&
0,4550&
0,0000 \\
\hline
0,5000&
0,0450&
0,0045&
0,4505 \\
\hline
0,0005&
0,4545&
0,4950&
0,0500 \\
\hline
\end{tabular}
\label{tab30}
\end{center}
\end{table}

\begin{itemize}
\item \textbf{Shannon entropy}
\end{itemize}
Based on above probability distributions, let us calculate the values of Shannon entropy. The table below give these values.
\begin{table}[htbp]
\begin{center}
\begin{tabular}{|l|l|l|l||l|}
\hline
&
&
&
&
0,3683 \\
\hline \hline
0,0400&
0,0114&
0,0610&
0,1506&
0,2630 \\
\hline
0,0646&
0,1511&
0,1556&
0,0000&
0,3713 \\
\hline
0,1505&
0,0606&
0,0106&
0,1560&
\textbf{0,3777} \\
\hline
0,0015&
0,1556&
0,1512&
0,0650&
0,3733 \\
\hline \hline
\textbf{0,2567}&
0,3788&
0,3783&
0,3716&
0,2667 \\
\hline
\end{tabular}
\label{tab31}
\end{center}
\end{table}

We observe that the value of Shannon entropy varies from 0,2567 to 0,3777.

\subsection{Shannon Entropy of Order 8x8}

In sections 3.2.1 and 3.2.2, we have two different kinds of magic squares of order $8 \times 8$. The magic square appearing in section 3.2.2 is bimagic. In both the cases, we considered here below the first one with binary digits. In these cases let us divide the each value by their magic sum. This gives us the following probability distributions:

\subsubsection{First case}

Here below is a table for probability distributions based on the binary magic square of order $8 \times 8$ given in section 3.2.1.
\begin{itemize}
\item\textbf{Probability distribution}
\end{itemize}
\begin{table}[htbp]
\begin{center}
\begin{tabular}{|l|l|l|l|l|l|l|l|}
\hline
0,00000&
0,22525&
0,24977&
0,02498&
0,00023&
0,22502&
0,25000&
0,02475 \\
\hline
0,24975&
0,02500&
0,00002&
0,22523&
0,24998&
0,02477&
0,00025&
0,22500 \\
\hline
0,02500&
0,24975&
0,22523&
0,00002&
0,02477&
0,24998&
0,22500&
0,00025 \\
\hline
0,22525&
0,00000&
0,02498&
0,24977&
0,22502&
0,00023&
0,02475&
0,25000 \\
\hline
0,00227&
0,22748&
0,24750&
0,02275&
0,00250&
0,22725&
0,24773&
0,02252 \\
\hline
0,24752&
0,02273&
0,00225&
0,22750&
0,24775&
0,02250&
0,00248&
0,22727 \\
\hline
0,02273&
0,24752&
0,22750&
0,00225&
0,02250&
0,24775&
0,22727&
0,00248 \\
\hline
0,22748&
0,00227&
0,02275&
0,24750&
0,22725&
0,00250&
0,02252&
0,24773 \\
\hline
\end{tabular}
\label{tab27}
\end{center}
\end{table}

\begin{itemize}
\item\textbf{Shannon entropy}
\end{itemize}

Based on above probability distributions, let us calculate the values of Shannon entropy. The table below give these values.
\begin{table}[htbp]
\begin{center}
\begin{tabular}{|l|l|l|l|l|l|l|l||l|}
\hline
&
&
&
&
&
&
&
&
0,6766 \\
\hline \hline
0,0000&
0,1458&
0,1505&
0,0400&
0,0008&
0,1458&
0,1505&
0,0398&
\textbf{0,6732} \\
\hline
0,1505&
0,0401&
0,0001&
0,1458&
0,1505&
0,0398&
0,0009&
0,1458&
0,6734 \\
\hline
0,0400&
0,1505&
0,1458&
0,0001&
0,0398&
0,1505&
0,1458&
0,0009&
0,6734 \\
\hline
0,1458&
0,0000&
0,0400&
0,1505&
0,1458&
0,0008&
0,0398&
0,1505&
0,6732 \\
\hline
0,0060&
0,1463&
0,1501&
0,0374&
0,0065&
0,1462&
0,1501&
0,0371&
\textbf{0,6797} \\
\hline
0,1501&
0,0373&
0,0060&
0,1463&
0,1501&
0,0371&
0,0065&
0,1462&
0,6796 \\
\hline
0,0374&
0,1501&
0,1463&
0,0060&
0,0371&
0,1501&
0,1462&
0,0065&
0,6796 \\
\hline
0,1463&
0,0060&
0,0374&
0,1501&
0,1462&
0,0065&
0,0371&
0,1501&
\textbf{0,6797} \\
\hline \hline
0,6761&
0,6761&
0,6761&
0,6761&
0,6768&
0,6768&
0,6769&
0,6769&
0,6763 \\
\hline
\end{tabular}
\label{tab28}
\end{center}
\end{table}

We observe that the values of Shannon entropy varies from 0,6732 to 0,6797

\subsubsection{Second case}

In this case we shall deal with bimagic square given in section 3.2.2 with binary digits. The table below is the probability distributions table:

\begin{itemize}
\item\textbf{Probability distribution}
\end{itemize}
\begin{table}[htbp]
\begin{center}
\begin{tabular}{|l|l|l|l|l|l|l|l|}
\hline
0,00250&
0,22725&
0,22502&
0,00023&
0,02477&
0,24998&
0,24775&
0,02250 \\
\hline
0,02475&
0,25000&
0,24773&
0,02252&
0,00248&
0,22727&
0,22500&
0,00025 \\
\hline
0,00000&
0,22525&
0,22748&
0,00227&
0,02273&
0,24752&
0,24975&
0,02500 \\
\hline
0,02275&
0,24750&
0,24977&
0,02498&
0,00002&
0,22523&
0,22750&
0,00225 \\
\hline
0,22523&
0,00002&
0,00225&
0,22750&
0,24750&
0,02275&
0,02498&
0,24977 \\
\hline
0,24752&
0,02273&
0,02500&
0,24975&
0,22525&
0,00000&
0,00227&
0,22748 \\
\hline
0,22727&
0,00248&
0,00025&
0,22500&
0,25000&
0,02475&
0,02252&
0,24773 \\
\hline
0,24998&
0,02477&
0,02250&
0,24775&
0,22725&
0,00250&
0,00023&
0,22502 \\
\hline
\end{tabular}
\label{tab29}
\end{center}
\end{table}

\begin{itemize}
\item \textbf{Shannon entropy}
\end{itemize}
\begin{table}[htbp]
\begin{center}
\begin{tabular}{|l|l|l|l|l|l|l|l||l|}
\hline
&
&
&
&
&
&
&
&
0,6763 \\
\hline \hline
0,0065&
0,1462&
0,1458&
0,0008&
0,0398&
0,1505&
0,1501&
0,0371&
0,6768 \\
\hline
0,0398&
0,1505&
0,1501&
0,0371&
0,0065&
0,1462&
0,1458&
0,0009&
\textbf{0,6769} \\
\hline
0,0000&
0,1458&
0,1463&
0,0060&
0,0374&
0,1501&
0,1505&
0,0400&
\textbf{0,6761} \\
\hline
0,0374&
0,1501&
0,1505&
0,0400&
0,0001&
0,1458&
0,1463&
0,0060&
\textbf{0,6761} \\
\hline
0,1458&
0,0001&
0,0060&
0,1463&
0,1501&
0,0374&
0,0400&
0,1505&
\textbf{0,6761} \\
\hline
0,1501&
0,0373&
0,0401&
0,1505&
0,1458&
0,0000&
0,0060&
0,1463&
\textbf{0,6761} \\
\hline
0,1462&
0,0065&
0,0009&
0,1458&
0,1505&
0,0398&
0,0371&
0,1501&
\textbf{0,6769} \\
\hline
0,1505&
0,0398&
0,0371&
0,1501&
0,1462&
0,0065&
0,0008&
0,1458&
0,6768 \\
\hline \hline
0,6763&
0,6763&
0,6766&
0,6766&
0,6764&
0,6763&
0,6766&
0,6766&
0,6763 \\
\hline
\end{tabular}
\label{tab30}
\end{center}
\end{table}

We observe that the values of Shannon entropy varies from 0,6761 to 0,6769.
Thus conclude that in the second case the variation is much less than in the
first case. Moreover in the second case the values are very much close to
each other. Since we know that magic square with probability distributions
is bimagic square. In this case we have $S2^{8\times 8}:=S(P)=0,2273$

\subsection{Shannon Entropy of Order 16x16}

Here we shall give directly the Shannon entropy table based on the first representation of bimagic square of order $16 \times 16$ given in section 4.
\begin{table}[htbp]\tiny
\begin{center}
\begin{tabular}{|l|l|l|l|l|l|l|l|l|l|l|l|l|l|l|l||l|}
\hline
&
&
&
&
&
&
&
&
&
&
&
&
&
&
&
&
0,97751 \\
\hline \hline
0,00000&
0,10681&
0,11283&
0,02377&
0,00045&
0,10675&
0,11289&
0,02360&
0,00335&
0,10738&
0,11230&
0,02211&
0,00363&
0,10732&
0,11235&
0,02194&
0,97749 \\
\hline
0,11283&
0,02379&
0,00006&
0,10680&
0,11288&
0,02362&
0,00049&
0,10675&
0,11230&
0,02210&
0,00332&
0,10738&
0,11236&
0,02192&
0,00360&
0,10733&
0,97752 \\
\hline
0,02379&
0,11283&
0,10680&
0,00006&
0,02362&
0,11288&
0,10675&
0,00048&
0,02209&
0,11230&
0,10739&
0,00332&
0,02193&
0,11236&
0,10733&
0,00360&
0,97752 \\
\hline
0,10681&
0,00001&
0,02377&
0,11283&
0,10675&
0,00044&
0,02361&
0,11289&
0,10738&
0,00335&
0,02211&
0,11230&
0,10732&
0,00363&
0,02194&
0,11235&
0,97749 \\
\hline
0,00335&
0,10738&
0,11230&
0,02211&
0,00363&
0,10732&
0,11235&
0,02194&
0,00001&
0,10681&
0,11283&
0,02377&
0,00044&
0,10675&
0,11289&
0,02361&
0,97749 \\
\hline
0,11230&
0,02209&
0,00332&
0,10739&
0,11236&
0,02193&
0,00360&
0,10733&
0,11283&
0,02379&
0,00006&
0,10680&
0,11288&
0,02362&
0,00048&
0,10675&
0,97752 \\
\hline
0,02210&
0,11230&
0,10738&
0,00332&
0,02192&
0,11236&
0,10733&
0,00360&
0,02379&
0,11283&
0,10680&
0,00006&
0,02362&
0,11288&
0,10675&
0,00049&
0,97752 \\
\hline
0,10738&
0,00335&
0,02211&
0,11230&
0,10732&
0,00363&
0,02194&
0,11235&
0,10681&
0,00000&
0,02377&
0,11283&
0,10675&
0,00045&
0,02360&
0,11289&
0,97749 \\
\hline
0,00360&
0,10733&
0,11236&
0,02192&
0,00332&
0,10738&
0,11230&
0,02210&
0,00049&
0,10675&
0,11288&
0,02362&
0,00006&
0,10680&
0,11283&
0,02379&
0,97752 \\
\hline
0,11235&
0,02194&
0,00363&
0,10732&
0,11230&
0,02211&
0,00335&
0,10738&
0,11289&
0,02360&
0,00045&
0,10675&
0,11283&
0,02377&
0,00000&
0,10681&
0,97749 \\
\hline
0,02194&
0,11235&
0,10732&
0,00363&
0,02211&
0,11230&
0,10738&
0,00335&
0,02361&
0,11289&
0,10675&
0,00044&
0,02377&
0,11283&
0,10681&
0,00001&
0,97749 \\
\hline
0,10733&
0,00360&
0,02193&
0,11236&
0,10739&
0,00332&
0,02209&
0,11230&
0,10675&
0,00048&
0,02362&
0,11288&
0,10680&
0,00006&
0,02379&
0,11283&
0,97752 \\
\hline
0,00048&
0,10675&
0,11288&
0,02362&
0,00006&
0,10680&
0,11283&
0,02379&
0,00360&
0,10733&
0,11236&
0,02193&
0,00332&
0,10739&
0,11230&
0,02209&
0,97752 \\
\hline
0,11289&
0,02361&
0,00044&
0,10675&
0,11283&
0,02377&
0,00001&
0,10681&
0,11235&
0,02194&
0,00363&
0,10732&
0,11230&
0,02211&
0,00335&
0,10738&
0,97749 \\
\hline
0,02360&
0,11289&
0,10675&
0,00045&
0,02377&
0,11283&
0,10681&
0,00000&
0,02194&
0,11235&
0,10732&
0,00363&
0,02211&
0,11230&
0,10738&
0,00335&
0,97749 \\
\hline
0,10675&
0,00049&
0,02362&
0,11288&
0,10680&
0,00006&
0,02379&
0,11283&
0,10733&
0,00360&
0,02192&
0,11236&
0,10738&
0,00332&
0,02210&
0,11230&
0,97752 \\
\hline \hline
0,97750&
0,97750&
0,97751&
0,97751&
0,97751&
0,97751&
0,97750&
0,97750&
0,97750&
0,97750&
0,97751&
0,97751&
0,97751&
0,97751&
0,97750&
0,97750&
0,97750 \\
\hline
\end{tabular}
\label{tab36}
\end{center}
\end{table}

In this case, the sum of the lines or columns are very much near to each other. Thus we
observe that in case of bimagic squares of order $8\times 8$ and $16\times
16$, sums representing Shannon entropy are very much close to each other in each case. In both the cases
the Shannon entropy is same upto three digit decimal. This gives that the bimagic squares give better results. 
The above magic square of order $16\times 16$ of probability distributions is bimagic square. In this case we have
$S2^{4\times 4}:=S(P)=0,11364$ This value is much more less than the value of Shannon's
entropy, i.e., approximately, $H(P)=0,9775$.

\begin{center}
------------------------
\end{center}

\end{document}